\documentclass[a4paper,fleqn]{cas-dc}

\usepackage[authoryear,longnamesfirst]{natbib}

\usepackage{amssymb}
\usepackage{amsmath}
\usepackage{xcolor}
\usepackage{hyperref}

\begin{document}
\let\WriteBookmarks\relax
\def\floatpagepagefraction{1}
\def\textpagefraction{.001}

\shorttitle{Statistical treatment of searches for counterparts of positionally-uncertain astrophysical sources}    

\shortauthors{Sitarek et al.}  



\title[mode = title]{Statistical treatment of searches for counterparts of positionally-uncertain astrophysical sources: from flux upper limits to detection.}

\author[inst1,inst2]{Julian Sitarek}[orcid=0000-0002-1659-5374]
\affiliation[inst1]{organization={Department of Astrophysics, University of Lodz},
             addressline={Pomorska 149},
             city={Lodz},
             postcode={90-236},
             country={Poland}}
\affiliation[inst2]{organization={Institut de Fisica d’Altes Energies },
            addressline={Campus UAB, Bellaterra},
            city={Barcelona},
            postcode={08193},
            country={Spain}}

\author[inst2]{Abelardo Moralejo}[orcid=0000-0002-1344-9080] 
\author[inst2]{Juan {Jiménez Quiles}}[orcid=0009-0005-6729-5709] 
\author[inst2]{Giacomo D'Amico}[orcid=0000-0001-6472-8381] 
\author[inst3]{Andrea Simongini}[orcid=0009-0000-3416-9865]
\author[inst3]{Antonio Stamerra}[orcid=0000-0002-9430-5264]
\affiliation[inst3]{organization={INAF - Osservatorio Astronomico di Roma},
            addressline={Via di Frascati 33},
            city={Monteporzio Catone},
            postcode={I-00078},
            country={Italy}}
\begin{abstract}
Rapid growth of the multimessenger and multiwavelength astrophysics had led to an increasing number of observations of the same events with instruments with different point spread functions. 
In particular pointing instruments with good angular resolution are used to pin-point the source of poorly-localized alert. 
In case no clear detection of the source counterpart is reached, the interpretation of the results requires statistical analysis. 
We investigate two approaches using the probability density function of the alert: frequentist and Bayesian, as well as agnostic approach not using this information.
We discuss the advantages and problems of all the methods, and compare their reliability and performance. 
We consider both a simple one-dimensional toy simulations and a realistic use case of full simulations of the follow-up of a gravitational wave event with gamma-ray telescopes. 
The performance of both frequentist and Bayesian approaches for weak signals is comparable and superior to the agnostic one.  
\end{abstract}


\begin{highlights}
\item We use frequentist and Bayesian approaches to exploit the prior knowledge on the alert position in the estimation of flux limits from a follow-up instrument. 
\item Both approaches result in comparable improvement of achieved constraints. 
\end{highlights}



\maketitle
\section{Introduction}

Over the past decade multimessenger astronomy grew into a fully-fledged field (see e.g. \citealp{2019NatRP...1..585M}) with a number of instruments devoted to observations of gamma rays, cosmic rays, neutrinos and gravitational waves (GW). 
Correlation of emission of different messengers became an important question in determining the nature of different types of sources \citep{2017ApJ...848L..12A, 2018Sci...361.1378I}.
Due to natural differences between detection techniques of various messengers it is common to achieve largely different localization accuracy with them. 
As a result, often poorly-localized alerts from e.g. GW detectors or some of neutrino channels are being followed up by superior angular resolution telescopes designed for detection of electromagnetic emission. 
Drawing conclusions about the source is complicated in the case no emission (or only hints of it) is seen. 

Different approaches of statistical analysis of such follow-up observations have been tried (see e.g. \citealp{2012PhRvD..85j3004B,2017ApJ...841L..16V,2019PhRvD.100h3017B,2019ICRC...36..918H,2020JCAP...05..016V}), aimed specifically at observations of different messengers (e.g. low rates of observed neutrino events requires careful treatment of Poissonian statistics in an unbinned analysis, while high rates of events observed in electromagnetic follow up might justify Gaussian approximation and binned treatment).  

In this work we review two types of methods: frequentist and Bayesian, which we describe in section~\ref{sec:met} and compare it to agnostic\footnote{While we call the method ``agnostic'' because it does not use the alert probability density values at different positions of the sky, the scan region itself is selected to cover the most-probable 95\% part of this distribution.} method that does not use the specific information about the probability density function of the alert position. 
In section~\ref{sec:toy} we present the basic features of those methods on a simple one-dimensional toy case and compare their performance. 
A realistic case of a GW follow-up with Cherenkov telescopes is discussed in section~\ref{sec:gw}.
We discuss the results in section~\ref{sec:conc}.
In appendixes we discuss it simplification of the calculations in the Bayesian method is possible (Appendix~\ref{sec:singlebin}) and discussion about including prior on positive flux (Appendix~\ref{sec:cl_f0}).
In Appendix~\ref{sec:syst} we investigating inclusion of systematic uncertainties in Bayesian method. 

\section{Scenario and investigated methods}\label{sec:met}
We consider an alert with a known source position probability density map, and a follow-up of this area with a scan by a second instrument. 
In the scenario that we investigate the data from the scanning instrument is burdened with Poissonian background (that can vary across the FoV of the instrument), and it is employed to detect potential emission from the source from which the alert originates. 
No other individual sources are considered\footnote{Such sources can be considered as part of the background model not altering the proposed framework.}.
Specifically, we do not address the association of newly detected sources with alerts when multiple candidates are present.

We assume that the dataset of the scanning instrument consists in the binned event counts in the scan region (i.e. a counts ``On-skymap''), where the individual events originate either from the background or from emission associated with the alert. 
Together with this ``On-skymap'' we have an analogous ``Off-skymap'' in which all recorded events are from the Poissonian background, and which is obtained with a larger exposure than that of the actual follow-up observation. 
We assume that the expected background rate in any given bin is the same for the On- and Off- skymaps. 
By setting  a finite exposure for the Off-skymap collection (e.g. 5$\times$ the On- exposure), we make the scenario realistic -- it is a way of representing our incomplete knowledge of the background.
This approach is also taking into account that background models are typically constructed from a dataset (either the same as the ``On-skymap'', or from a separate one). 
For all the purposes of the present study, this is equivalent to having a background model with uncertainties (although the specific implementation of the algorithms would slightly change depending on the details of the model uncertainties). 
Using a very large ``Off-skymap'' exposure would correspond to having a background model with negligible uncertainty.

We consider three methods, representing three different approaches. 
We discuss the calculation of the statistical significance (over the background fluctuations) of the possible detection of coincident emission  by the scanning instrument.
For each method we show how the upper limit on the emitted flux can be computed in case of no significant detection. 
We do not focus on the question of evaluating the flux when the detection is clear, because in such a case the positional uncertainty becomes small and standard procedures for sources of known location can be applied.
To simplify the comparison between the methods we 
report throughout the paper the confidence level of the upper limits computed relative to the part of the alert probability map covered by the scan. 
In a realistic case in which only limited part of the probability density map is covered, the confidence level  of the final limit can be computed by multiplication of the confidence level used in the calculation with the coverage fraction of the probability map. 
Notably, this also means that a scan covering a sizeable fraction of the alert uncertainty region, but far from 100\%, is justified only as an attempt to detect a counterpart, as it will not provide meaningful constraints to the emission, in case no significant signal is found.

The agnostic method consists of taking the least-constraining upper limits from the whole scanned region.
The frequentist approach is based on the method of \cite{2019ICRC...36..918H}, 
however converted into a binned version.
The Bayesian approach uses the classical Bayesian posterior derivations using alert probability map as the prior on the source position.  

\subsection{Agnostic method}
At each position within the scan region one can derive not only the significance of a possible detection, but also an upper limit on the flux from this location, creating a map of upper limits (see e.g. \citealp{2025arXiv251216562A}). 
The commonly used way of computing such limits is following the (frequentist, known-location) approach described in \cite{2005NIMPA.551..493R}.
We use the unbounded implementation of the method \citep{2010CoPhC.181..683L} available in the ROOT data analysis package \citep{refroot}. 
In the absence of a clear signal and precise source location, one can take the least constraining upper limit from the part of the map covering the alert location as the flux upper limit for a source within the scanned region. 
In a realistic case, the large range of possible source positions also results in a large range of statistical fluctuations in the number of events at various positions, even if they have similar exposure. 
Such fluctuations produce a large spread of the individual upper limits.
For example, if a region of 10 deg$^2$ is scanned with an instrument with angular resolution of $\sim0.1^\circ$, the number of ``trials'' in which the signal can be searched is $\sim10/(\pi 0.1^2)=320$, and spurious fluctuations up to $\sim2.7\sigma$ above the background are to be expected. 
Notably, the neighbouring positions in the skymap might have partial correlation, however it is not straightforward to include this in the calculation of the number of trials, hence normally those correlations are ignored resulting in a conservative (maximal) number of trials considered. 
Assuming large statistics ($\gg 10$ events) and negligible systematic uncertainties a 95\% C.L. (confidence level) limit for such a 2.7$\sigma$ excess is $\sim2.6$ times larger than for a $\sim0\sigma$ (no excess) case. 
Additionally, possible lower exposure observations at the edge of the scan region can further degrade the agnostic limit, even if the corresponding probability of the alert originating from there is low.

\subsection{Frequentist approach}\label{sec:freq}
An improved frequentist approach can be implemented, profiting from the values of the probability density map of the alert. We refer to this as the ``Frequentist approach'' for simplicity. We adopt the method of \cite{2019ICRC...36..918H}.
However, we adapt it to the high rate of events expected from Cherenkov telescopes by considering the binned implementation of the method. 
Using an event-wise, unbinned method in such a case would be computationally prohibitive and is not expected to bring any improvement.  
We construct a test statistic for each considered source position $x$:
\begin{equation}
   \mathrm{TS}(x)=2 \ln(L_{src}(x,\hat{\mu}, \hat{b})/L_0(x,\hat{b_0})) + 2 \ln(p_{GW}(x)), \label{eq:freq}
\end{equation}
where
$L_{src}$ is the likelihood computed under the assumption of the presence of the source, $L_0$ is the likelihood of the null hypothesis (all recorded events are background) and $p_{GW}$ is the probability density of the alert position. 

The likelihoods are computed using the events recorded around the position x in the On- and Off skymaps (this is why $L_0$ depends on x). In case the signal is expected to spread beyond the bin containing the source (in the On-skymap), the event statistics $N_{ON}$, $N_{OFF}$ used in the calculation can be obtained by integration in a region around x, the optimal size of which depends (for point-like sources) on the instrument's angular resolution.

The estimated values of the signal and background, $\hat{\mu}$, $\hat{b}$ and $\hat{b_0}$ are, for each $x$, those which maximize 
the likelihoods of the two hypotheses given the data.

The likelihood ratio $L_{src}(x,\hat{\mu}, \hat{b})/L_0(x,\hat{b_0})$ can be computed following Eq.~14 of \citet{1983ApJ...272..317L}.
Notably, the above formula would provide high TS values both for positive and (unphysical) negative fluxes. 
As we aim to use it for detecting excesses, we set $L_{src}(x,\hat{\mu}, \hat{b})/L_0(x,\hat{b_0})$ to one if $\hat{\mu}<0$.

Next, the TS value is maximized over position $x$: $\mathrm{TS}=\max_{x} \mathrm{TS(x)}$.
In order to determine the significance associated to a given observed value, TS$^*$, we construct via Monte Carlo simulations the distribution of TS in the background-only scenario. The chance probability of the observation is then computed as the fraction of realizations with $\mathrm{TS}>\mathrm{TS^*}$.
In case of lack of detection, similar MC simulations are done together with the signal corresponding to a given flux injected randomly following the $p_{GW}$ probability distribution. 
The value of the upper limit is selected such that the fraction of obtained TS values larger than TS$^*$ is equal to the requested C.L. of the limits.

This causes a significant caveat if the TS value obtained in a given realization is far in the low values tail, in particular below the $1-$C.L. quantile.   
Then, by the above definition, the C.L. of the zero flux limit would be already above the requested C.L., and nominally a non-physical negative flux limit would be obtained.
To counteract this, we clip the obtained TS values to be at least equal to the median of the TS distribution obtained for the null hypothesis. An alternative solution to this problem is discussed in Appendix~\ref{sec:cl_f0}.
Clipping TS to its median for the null hypothesis is a conservative approach: we do not {\it profit} even from moderate down-fluctuations which would result in a positive (hence physical) upper limit. With such a procedure one would never obtain an artificially too good upper limit as a result of a systematic error, should that be the reason for the low TS value. 

We note that the above-described method is only one of the possibilities of the large family of frequentist approaches. 
Various methods can be constructed by taking different forms of Eq.~\ref{eq:freq}. 
In particular, the angular resolution of the instrument can be incorporated in the likelihood through the use of its point-spread function to obtain the expected number of signal counts in each bin of the On-skymap, and building a likelihood with one Poissonian term per bin (for bins around the assumed source position x, or in the whole skymap).

As the basic frequentist approach used in the next sections we apply the same scheme as in the agnostic method, namely integrating the counts up to a given cut value. 

\subsection{Bayesian method}\label{sec:bayes}
We consider the probability distribution of the position, $x$, and flux, $f$, of the source independent from each other.   
To determine the probability distribution of the flux of the source we start from the general Bayesian formula.
We apply the alert probability distribution, $p_{GW}$, as the prior for the position and a flat prior on the flux.  
The posterior probability distribution is computed as a multiplication of probabilities of all the bins $x'$ of the scan:
\begin{eqnarray}
    p(x, f)&=&A\prod_{x'}(P(N_{ON}(x'), \mathrm{PSF}(x,x')\mu+b(x')) \nonumber \\
    &\times& P(N_{OFF}(x'), b(x')\beta)\times p_{GW}(x')),  \label{eq:pbayes}  
\end{eqnarray}
where $P(k, \lambda)=\lambda^ke^{-\lambda}/k!$ is the Poissonian probability mass function. 
$N_{OFF}(x')$ is the number of measured background events at position $x'$.
However, is it straightforward to generalize the formula to any background model. 
In particular, for perfectly known expected background (i.e., a model with no uncertainty), corresponding to $\beta\to\infty$, one can skip the $P(N_{OFF}(x'), b(x')\beta)$ factor in Eq.~\ref{eq:pbayes}).
The factor $P(N_{OFF}(x'), b(x')\beta)$ can be also substituted with a Gaussian factor describing the uncertainty of the background at different positions of the skymap.

Notably, the multiplication of probabilities is done over all the bins.
Limiting $x'$ only to a region of interest around the $x$ position, as done by \cite{2017ApJ...841L..16V}, while significantly simplifying the computations, would have side effects that we discuss in Appendix~\ref{sec:singlebin}.

The background parameter $b(x')$ that maximizes the likelihood can be computed analytically \citep{2005NIMPA.551..493R} from the event statistics in the ON and OFF regions, for any assumed value of the signal parameter $\mu$:
\begin{eqnarray}
\Delta(x')&=&(\mu(1+\beta) - N_{ON}(x')-N_{OFF}(x'))^2 \\\nonumber
&+& 4 (\beta+1)\mu N_{OFF}(x') \\
b(x')&=&\frac{\sqrt{\Delta(x')}\!-\!\mu(1\!+\!\beta)\!+\!N_{ON}(x')\!+\!N_{OFF}(x')}{2(\beta+1)}\qquad
\end{eqnarray}
The normalization constant $A$ is obtained from normalizing  $\sum_{x} \int p(x,f)\; \mathrm{d}f = 1$, and $\mathrm{PSF}(x,x')$ is defined as the fraction of the signal originated at position $x$ being reconstructed at $x'$.
While $x'$ follows the binning of the data, for simplicity we also assume that $x$ can be considered as discrete values (sufficiently finely binned comparing to the PSF of the instrument). 
The probability distribution can be marginalized over position:
\begin{equation}
    p_f(f) = \sum_x p(x,f).\label{eq:pf}
\end{equation}
Knowing $p_f(f)$ it is straightforward to compute an upper limit $f_{UL}$ on the emission with a given C.L. from the condition:
\begin{equation}
    \mathrm{C.L.}=\int_0^{f_{UL}} p_f(f)\; \mathrm{d}f \label{eq:cl}
\end{equation}

Similarly to Eq.~\ref{eq:cl}, the estimation of the flux of the source, as well as its $1\sigma$ uncertainties, can be obtained by computing the median and 16\% and 84\% quantiles of the probability distribution.

Notably, contrary to the method presented in Section~\ref{sec:freq}, the Bayesian approach considers in a natural way only positive fluxes. 
Therefore, negative fluctuations never result in unphysical (negative) upper limits, and can be fed to the method without any problem (unless, of course, one suspects the down-fluctuation is not statistical, but the result of hidden systematic errors). 

\section{Toy case}\label{sec:toy}
\begin{figure}
    \centering
    \includegraphics[width=0.99\linewidth]{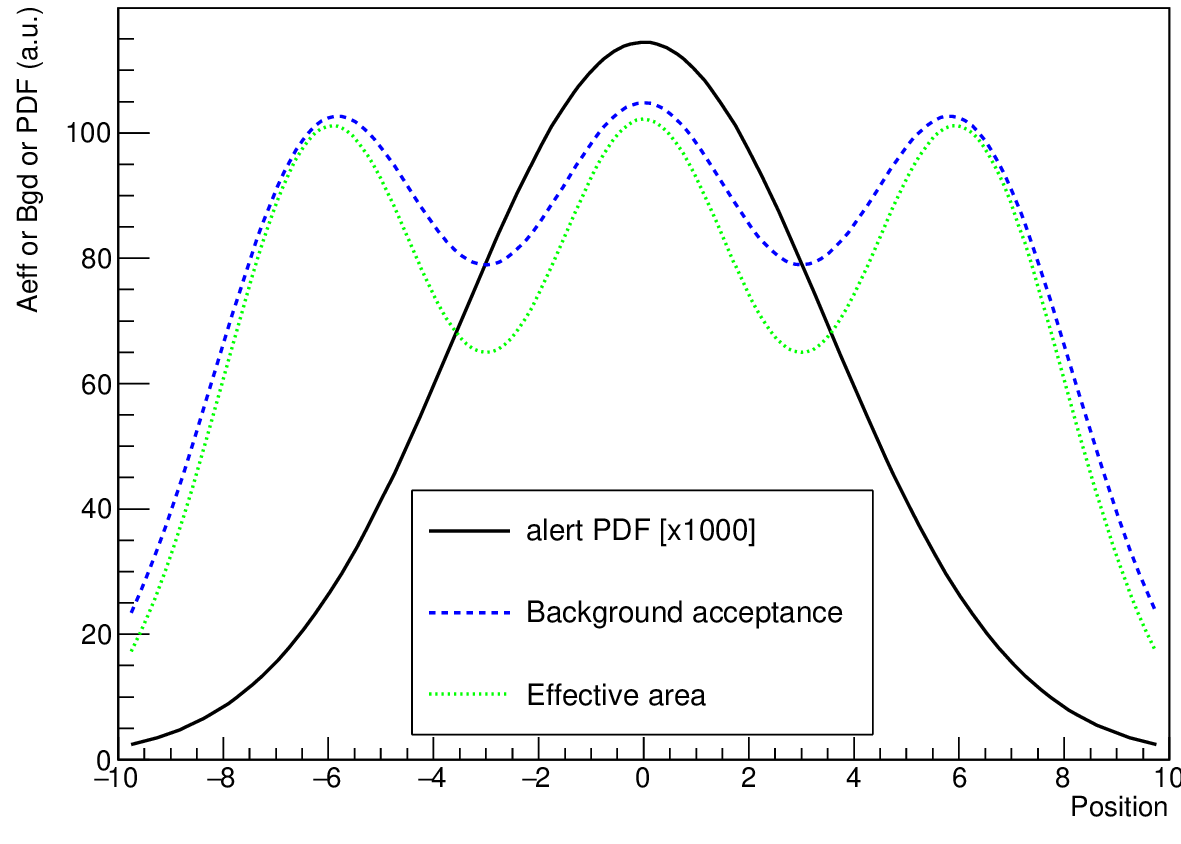}
    \caption{Toy MC setup. Solid black line shows $p_{GW}(x)$, the alert PDF (upscaled by a factor of 1000), blue dashed line shows $b(x)$, the background acceptance (in the units of counts) green dotted line shows the effective area (in the units of counts per unit flux).}
    \label{fig:toy_setup}
\end{figure}

In order to compare the basic features of the methods, we first apply them on a toy MC case. 
We assume a single position parameter, $x$, divided into $N_b=40$ bins. 
We further assume perfect reconstruction of the signal, i.e. $\mathrm{PSF}(x,x')=\delta(x-x')$, where $\delta$ is Dirac's delta function. 
Notably, in this way the proposed frequentist, Bayesian and agnostic approaches can be compared fairly without considering possible gain of using PSF information in the method.
The probability distribution function of the position of the alert is modelled as a Gaussian with clipped tails.
To facilitate the comparison we assume that the whole range of possible alert locations is covered by the scan.
The scan exposure is modelled as a combination of three Gaussians (with a slightly different width for the signal and background acceptance, see Fig.~\ref{fig:toy_setup}), to mimic a typical situation of a few separate pointings. 
Each iteration of the simulation consists of randomizing the source position drawing it from the $p_{GW}$ distribution. 
Next, in each of the $N_b$ bins $N_{OFF}(x)$ is computed by drawing from the Poisson distribution with mean of $\beta b$. 
$N_{ON}(x)$ is computed as a sum of a number drawn from a Poissonian with mean of $b$ (background component), and in case $x$ matches the source position, a second Poissonian component with mean $\mu$ equal to the assumed flux times effective area is added (signal component).   
Then, each of the methods described in Section~\ref{sec:met} is applied to the same simulated dataset to compute a 95\% C.L. upper limit on the emission. 

\begin{figure*}
\centering
    \includegraphics[width=0.3\textwidth]{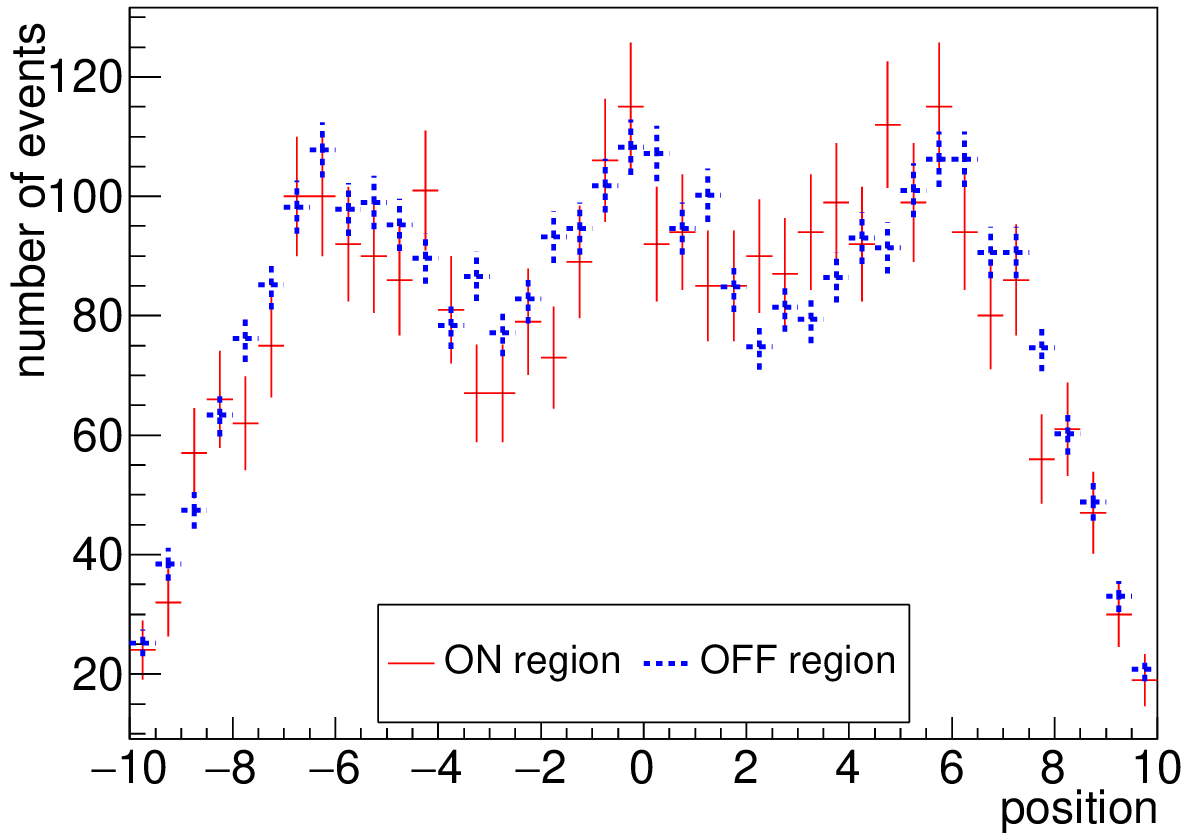}
    \includegraphics[width=0.3\textwidth]{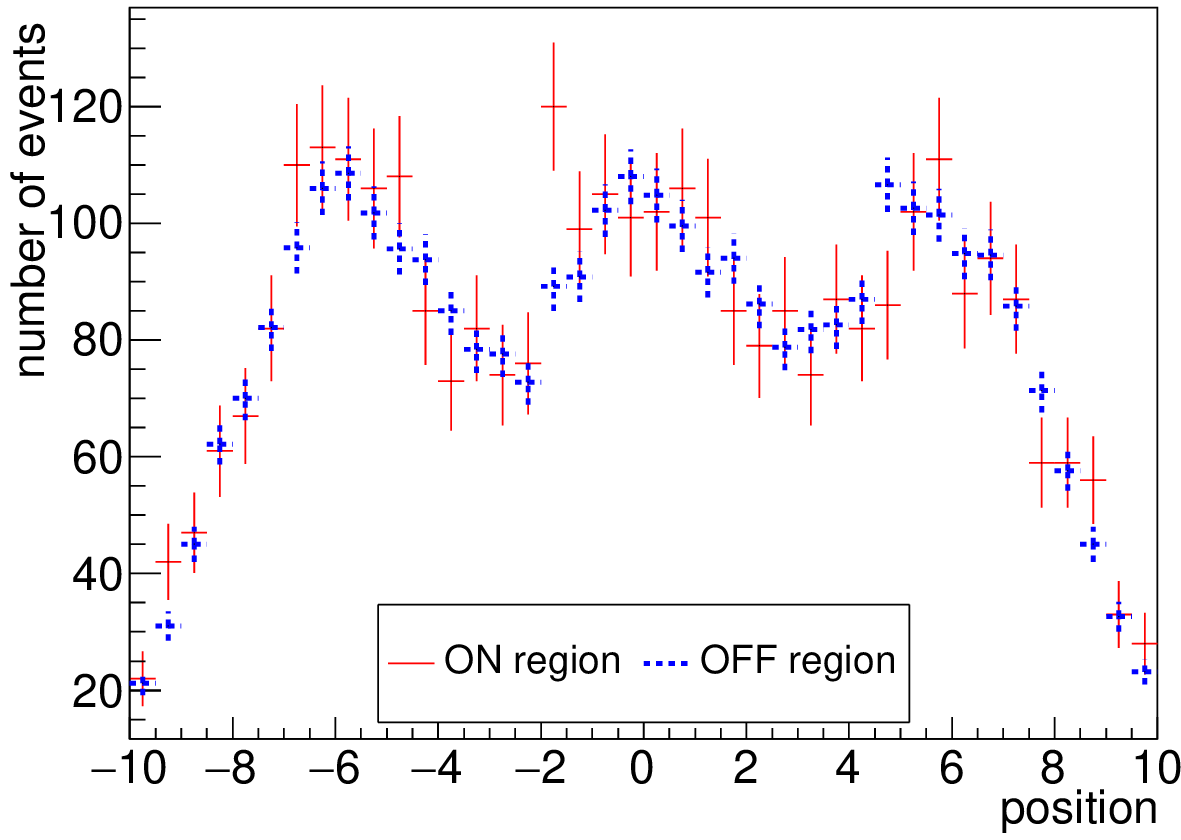}
    \includegraphics[width=0.3\textwidth]{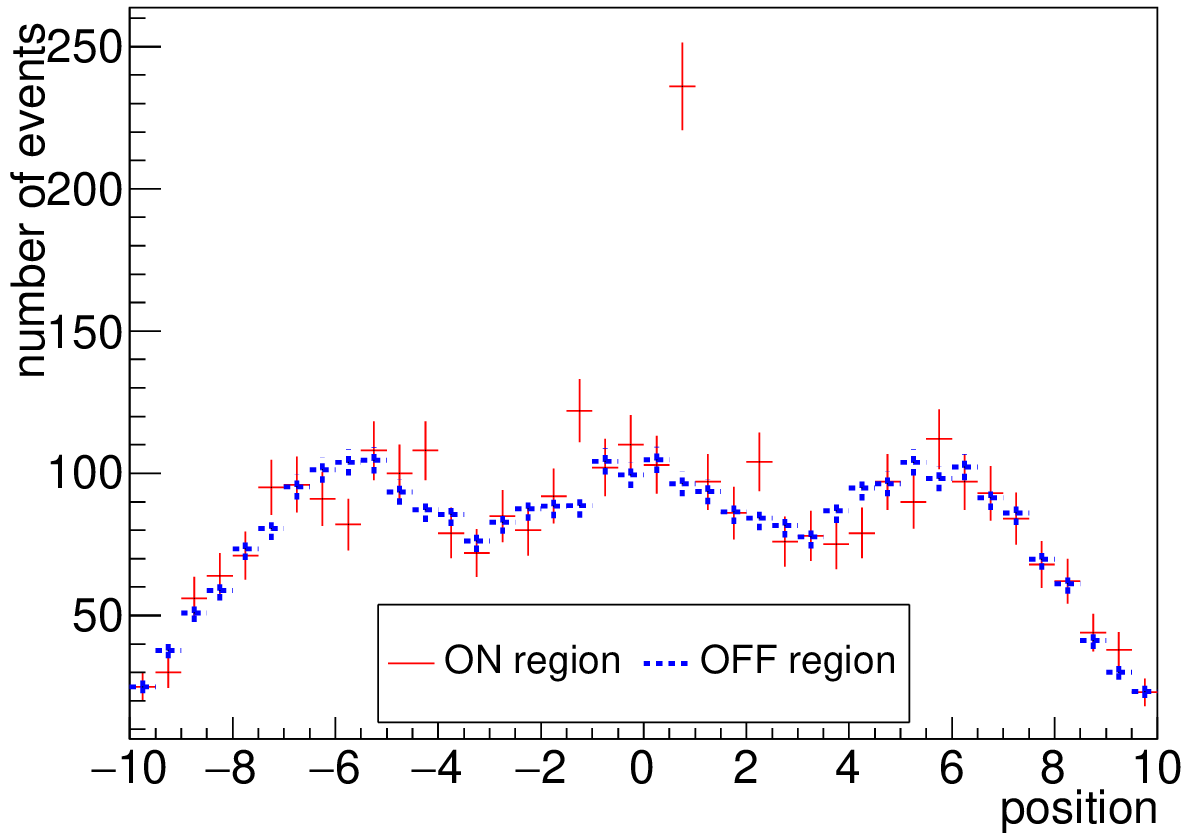}
    \includegraphics[width=0.3\textwidth]{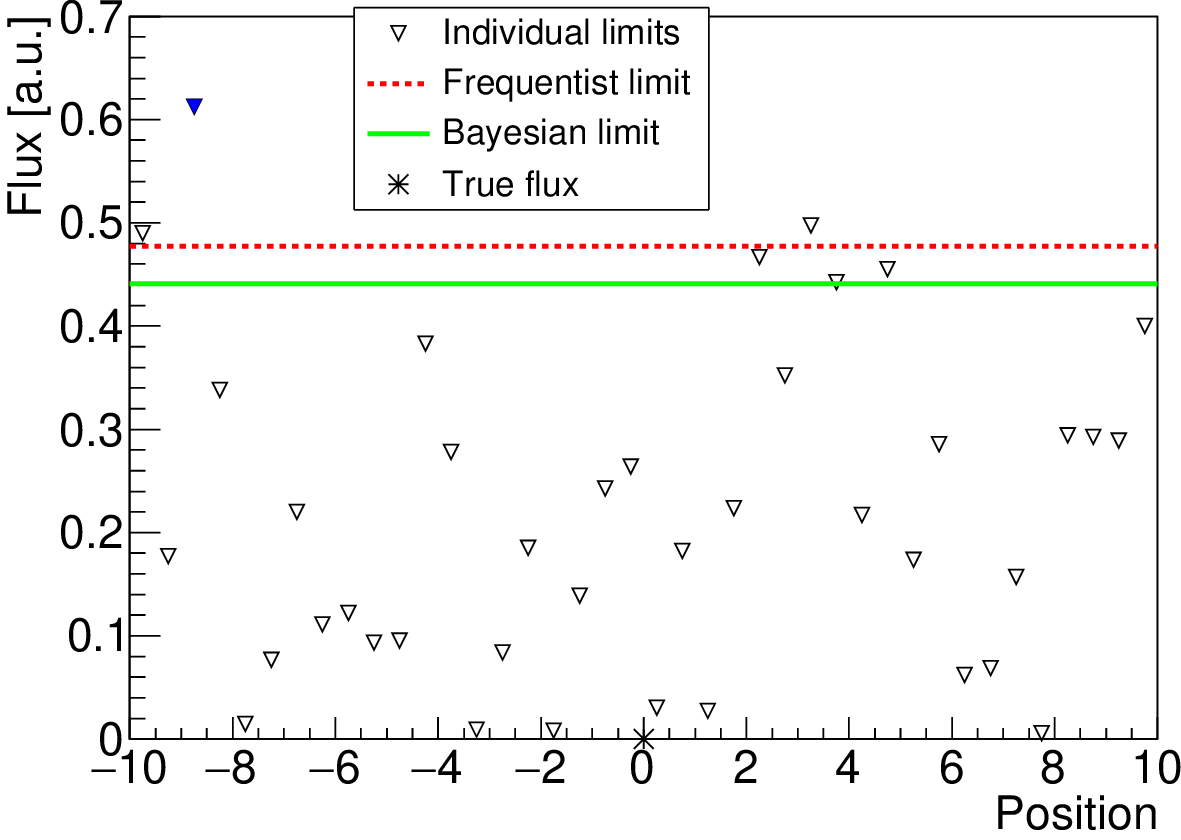}
    \includegraphics[width=0.3\textwidth]{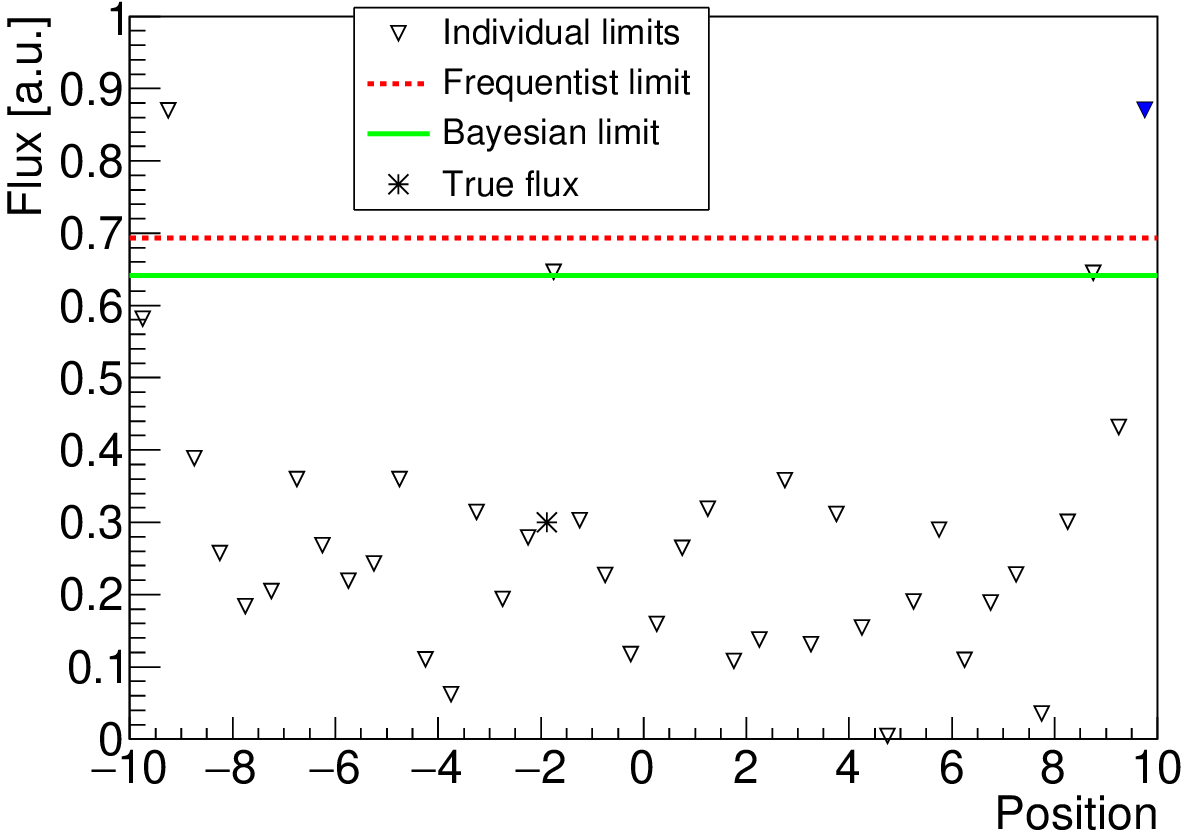}
    \includegraphics[width=0.3\textwidth]{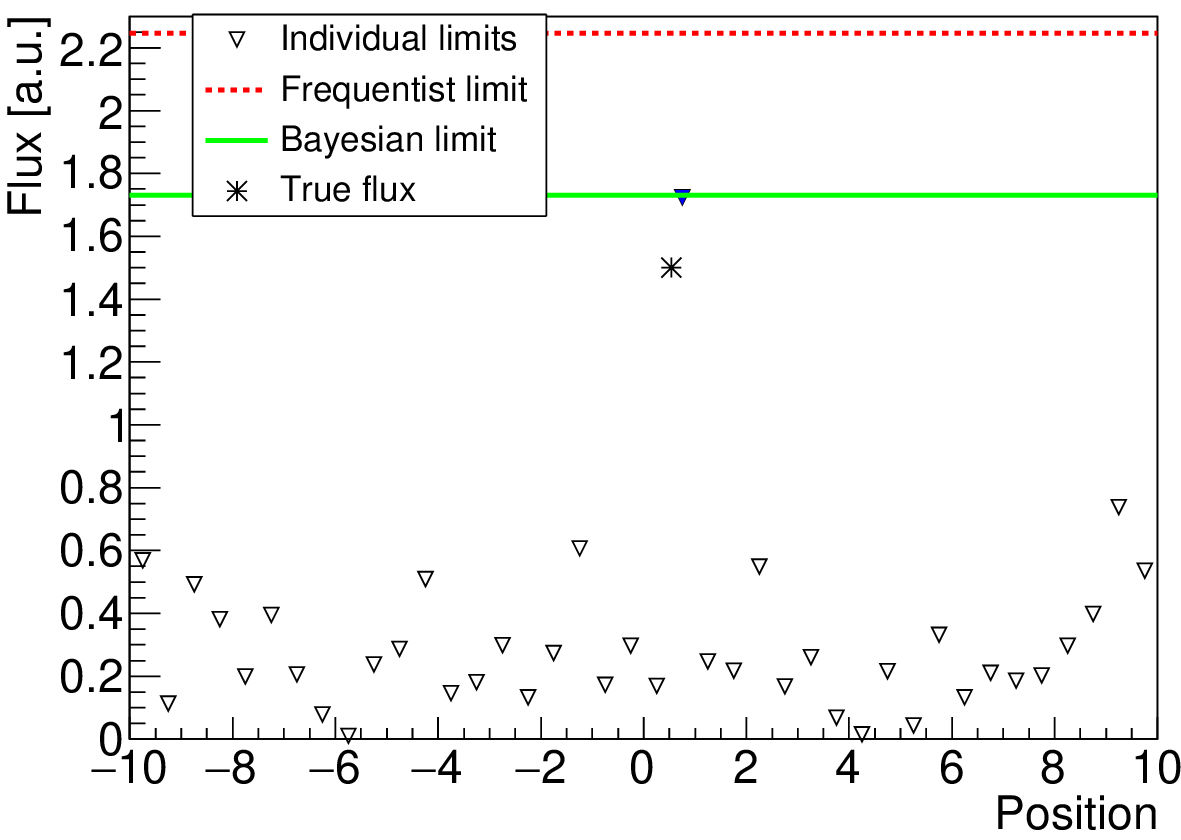}
    \caption{Examples of the three methods applied to a single toy MC realization with a true flux of 0 (left panels), 0.3 (middle panels) and 1.5 (right panels). 
    Top panels show the number of counts as a function of the position in ON region (solid red) and the corresponding background (OFF region, dotted blue).
    Bottom panels show the limits at individual positions using \cite{2005NIMPA.551..493R} (empty downward triangles, with the least constraining limit shown as filled marker) and global limits according to frequentist (red dashed) and Bayesian (green solid) approach.
    True flux (and its position) is shown with an asterisk.} 
    \label{fig:toy_examples}
\end{figure*}

The three example realizations (for no signal, weak signal and strong signal) are presented in Fig.~\ref{fig:toy_examples}. 
The scatter of the limits at individual positions due to varying event statistics is clearly seen, as well as less restrictive limits at the edge of the scan region where the exposure is lower.
Both effects result in the agnostic limit not being very strong.
In case of signal, the corresponding upper limit is visibly larger at the source position to accommodate the excess.

\begin{figure*}
    \includegraphics[width=0.95\textwidth]{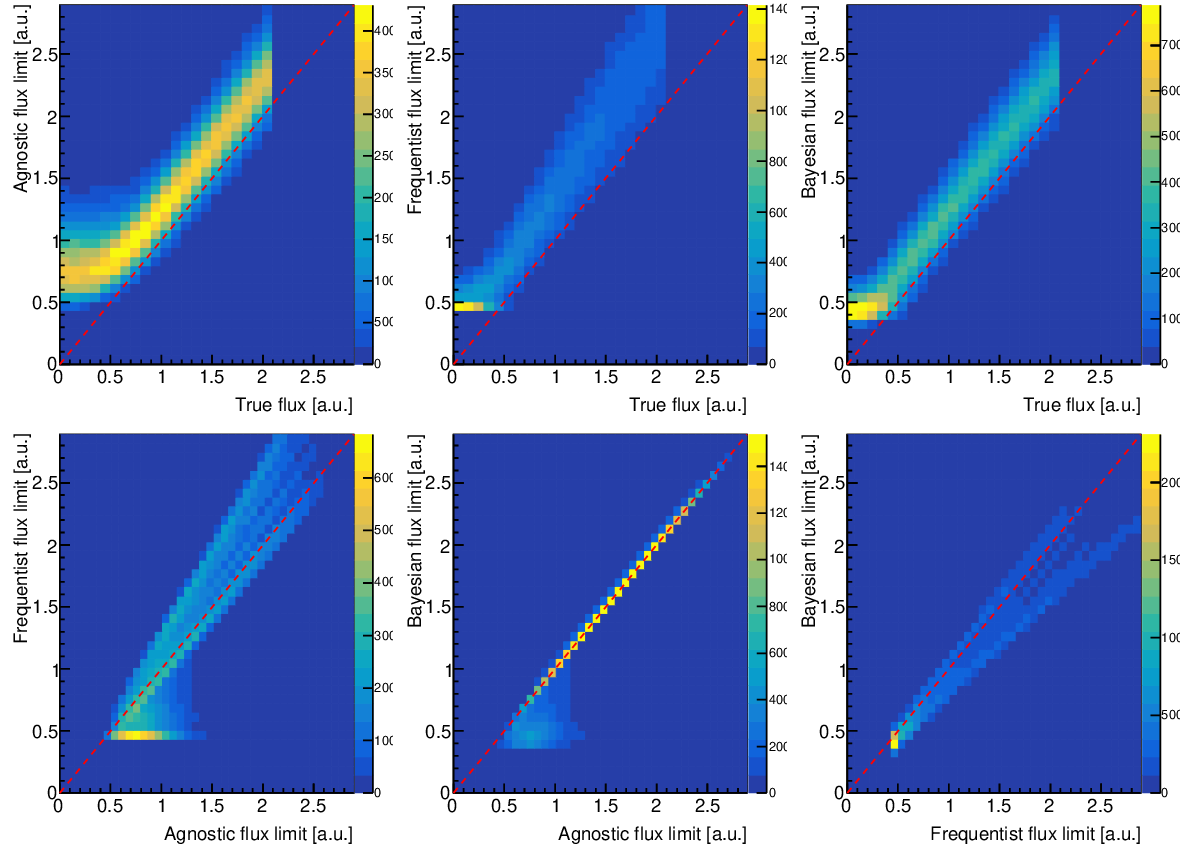}
    \caption{Comparison of the performance of the three methods in deriving upper limits on the flux in toy MC simulations. Top panels: limits obtained with different method as a function of the true flux for agnostic (top left), frequentist (top middle) and Bayesian (top right) method.
    Bottom panels show comparison of limits obtained with the three methods: frequenstist vs agnostic (bottom left), Bayesian vs agnostic (bottom middle) and Bayesian vs frequentist (bottom right). }
    \label{fig:toycomp}
\end{figure*}

\begin{figure}
    \centering
    \includegraphics[width=0.45\textwidth]{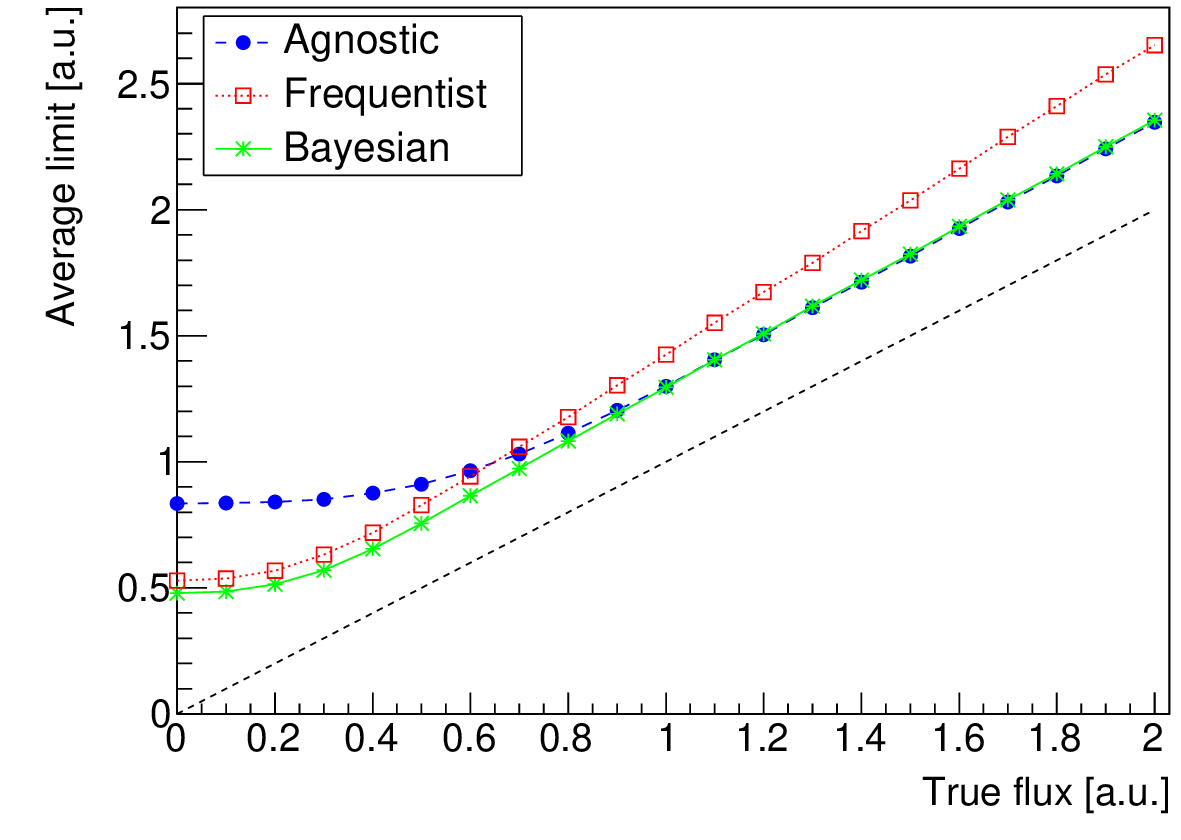}
    \caption{Comparison of the average limit obtained from many realizations of the same simulated true flux for agnostic (blue filled circles and dashed line), frequentist (red empty square and dotted line) and Bayesian (green asterisks and solid line). }
    \label{fig:toy_avrlimit}
\end{figure}

All three methods are compared in Fig.~\ref{fig:toycomp} for a population of realizations with different true flux assumed. 
The average upper limits obtained by each method for different realizations of the same true flux value are compared in Fig.~\ref{fig:toy_avrlimit}

Each realization has a randomly selected position of the source according to the $p_{GW}$ distribution, and event statistics fluctuated in a Poissonian way. 
As the flux approaches 0 all methods converge to their corresponding distribution of upper limits for null flux. 
The limit values in this case are comparable between the frequentist and Bayesian approach, while the agnostic method is significantly worse.
In the case of a strong signal, the Bayesian method converges to the same value as the agnostic one. This is understandable, as in this case the prior on the source position is very weak compared to the posterior obtained from the source detection. 
Therefore, the Bayesian method converges into the regular known-position point source analysis, and so does the agnostic method: for a strong flux, the selected, less-constraining upper limit will always be the one obtained at the source position. The frequentist upper limits, on the other hand, are higher and show a wider spread for the case of a strong source: this is understandable, since the toy MC realizations to obtain the limit are generated over the whole PDF of the alert location (regardless of the position of the actually observed excess). This includes regions of different exposure, where a given flux produces a different TS distribution, and in turn this broadens the global TS distribution from which the target coverage is computed.
In practical situations it is not a problem, as for such a strong detection, regular point-like analysis can be used to derive the confidence interval of the flux.

\begin{figure}
    \centering
    \includegraphics[width=0.45\textwidth]{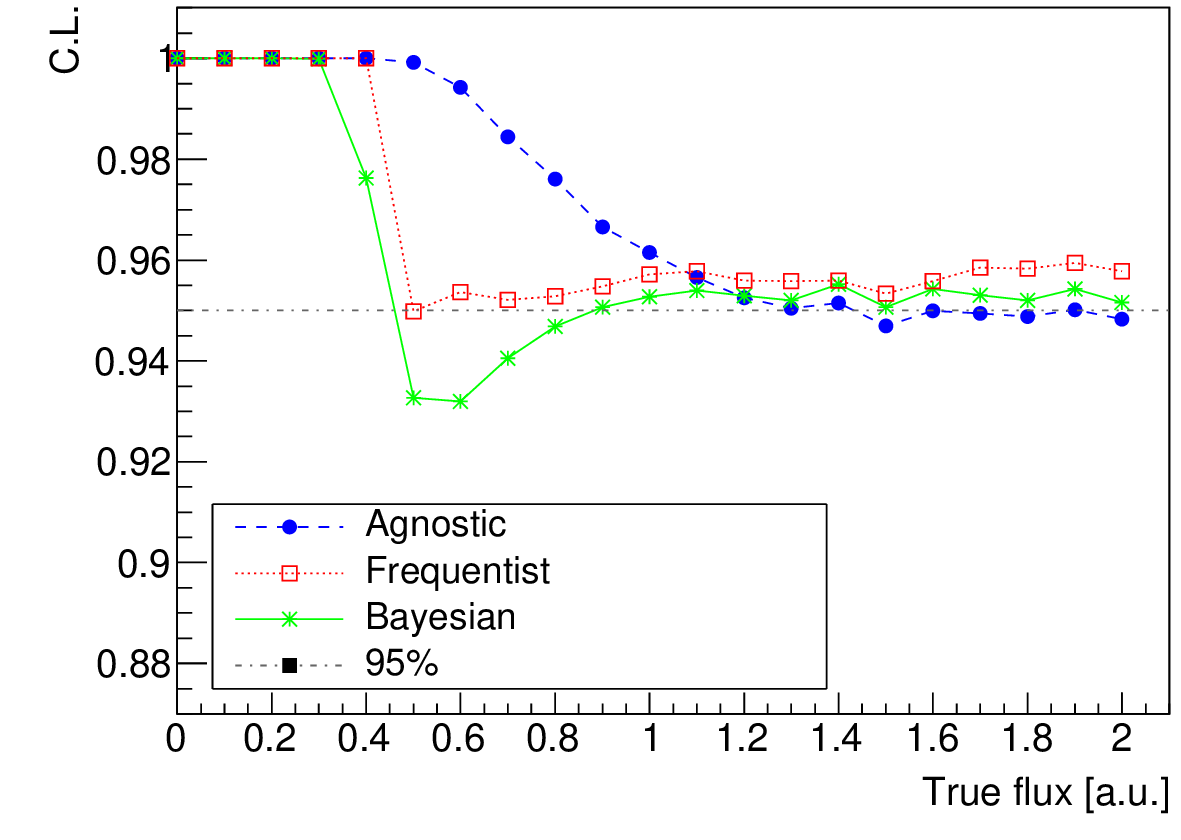}
    \caption{Fraction of realization of toy MC with a given true flux that result in the limit above the true flux value for the three methods: agnostic (blue filled circles and dashed line), frequentist (red empty square and dotted line) and Bayesian (green asterisks and solid line). 95\% is shown with gray dot-dashed line.
    }
    \label{fig:toy_cl}
\end{figure}

Finally, we compute the coverage of the limits for a case of a particular simulated flux. 
In Fig.~\ref{fig:toy_cl} we present what is the fraction of the limits above a particular value of the true flux.  
Since in all the methods we consider only positive flux limits, for a close to zero true flux this coverage approaches 100\%.
For large values of the true flux the frequentist approach results in C.L. close to the requested 95\%. 
For the lowest true fluxes where the extra condition of using the median of the distribution of the test statistics for TS downfluctuations makes the C.L. also approach 100\%). 
For the Bayesian method the empirical coverage drops slightly below the nominal 95\%, down to $\sim93\%$ at the true flux of $\sim$0.5~a.u.
At such flux values the excess is expected to be large enough to provide a hint of signal with chance probability comparable with the C.L. (see next section).
The agnostic method is very conservative and providing poor limits in this range, which have the empirical C.L. considerably above the 95\%. 
At higher fluxes, when the signal starts to be highly significant, the Bayesian method provides much stronger posterior on the source position compared to the prior. 
As a result the obtained limit is the same as in agnostic case (see the explanation of Fig.~\ref{fig:toy_examples}c), and the empirical C.L. is the same for both methods and equal to the requested C.L. of 95\%. 

While both Bayesian and frequentist methods provide an answer that can be interpreted as a limit on the observed flux, the exact question that they answer is different. 
Namely, frequentist method is looking for a flux that provides requested coverage of the upper limit, i.e. comparing the same flux, but different realizations of event statistics. 
This is exactly the same quantity as verified in Fig.~\ref{fig:toy_cl}, thus it is expected to have a very good agreement with the requested C.L. for this method.
On the other hand, the Bayesian method is finding a posterior probability for a given flux distribution prior assuming the same statistics as measured in a given realization (i.e. various fluxes but the same event statistics). 
Therefore, for the Bayesian method it is not given for granted that the achieved ``frequentist-style'' coverage of the limits matches the requested one. 
However, as we show in Fig.~\ref{fig:toy_cl} the deviation from the requested C.L. is rather small. 
It is possible to imagine an opposite ``Bayesian-style'' test: taking a base realization, then generating a huge amount of random realizations with a flux following a prior but selecting only those that exactly match the event statistics in the base realization.
Over those selected realizations the 1-C.L. quantile of the selected fluxes could be then compared with the limit obtained from the base realization.
In such a test it is expected that the Bayesian method would provide perfect results (down to numerical accuracy of the calculations), while frequentist limit will not provide exactly the requested C.L.. 
Unfortunately, it is not possible to perform this kind of test even in this simplified toy MC case, as randomly matching exact statistics of all the signal and background bins is extremely improbable.

\subsection{Detection chance probability}\label{sec:detection}
We investigated the power of all the three methods in detecting a source. 
We simulated various flux states and calculated ``significance'' measures of each of the methods using the toy MC approach.
For each flux level, we take the median value of such a measure and compute the corresponding probability of reaching such a value with fluctuations of pure background. 

\begin{figure}
    \centering
    \includegraphics[width=0.45\textwidth]{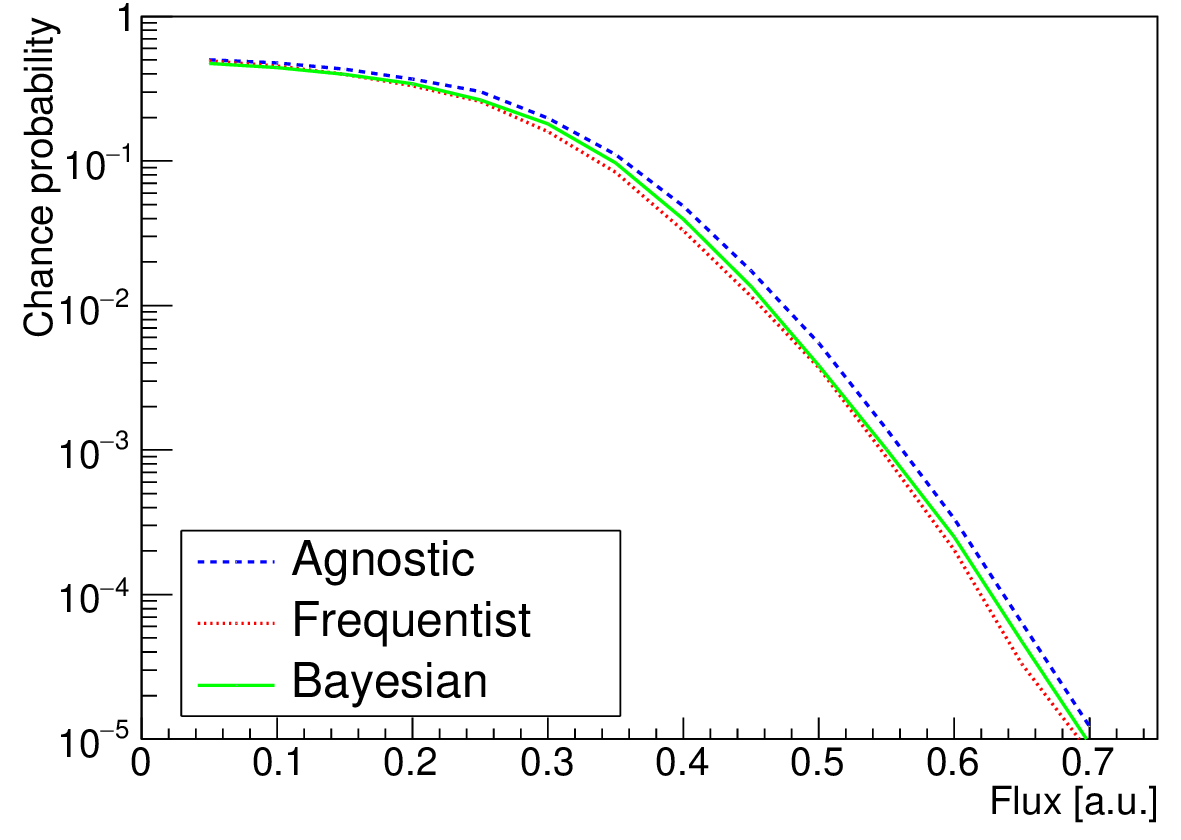}    
    \caption{The chance probability of observing, from background alone, a fluctuation at least as large as the observed excess in agnostic (blue dashed), frequentist (red dotted ) and Bayesian ( green solid) approach using toy MC simulations.}
    \label{fig:detect_toy}
\end{figure}

In the case of the agnostic method, we use the classical \citet{1983ApJ...272..317L} formula to calculate the significance in each of the possible positions. 
The significance is maximized over all the bins and is analytically converted to a one-sided chance probability (of the background to produce such an excess, or a larger one) by integrating the Gaussian distribution above the given significance value. 
Next, such $p_1$ probability is corrected by $n_x$ trials (connected to the number of possible positions) with the $1-(1-p_1)^{n_x}$ formula. 
The $n_x$ factor is not straightforward to compute as the nearby positions in the skymap are not fully independent due to integration of event statistics around each position. 
Here we conservatively assume that it is equal to the number of investigated true source positions. 
The resulting chance probability is shown in Fig.~\ref{fig:detect_toy}.

In the case of the frequentist method, we use the distribution of TS values computed according to Eq.~\ref{eq:freq}. 
For each flux level we determine the median TS value, and then compute the corresponding quantile of the distribution for null flux.

\begin{figure}
    \centering
    \includegraphics[width=0.45\textwidth]{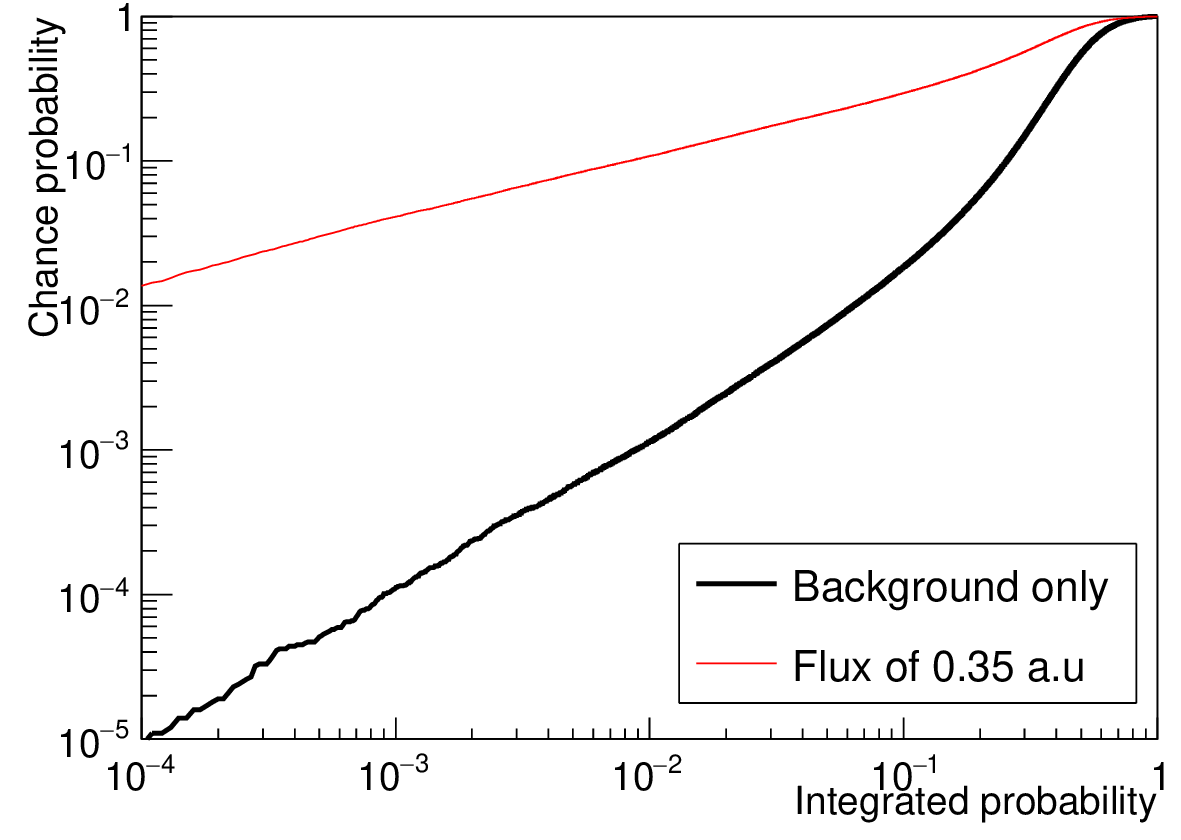}
    \caption{Cumulative distribution of the posterior probability of ``negative'' flux for a simulated null flux (thick black line) and 0.5 a.u. flux (thin red line) for toy MC simulations.}
    \label{fig:bayescalib}
\end{figure}

In the case of the Bayesian method, there is no direct equivalent of frequentist significance. 
Moreover, as the flux is a continuous quantity, the derived posterior probability of exactly zero value is also equal to zero. 
In order to avoid this, we can artificially allow non-physical negative flux values in the calculation of the probability distribution and integrate the marginalized probability $p_0=\int_{-\infty}^0 p_f(f)\; \mathrm{d}f$.
The $p_0$ distribution for the null hypothesis is not uniform, which means $p_0$ cannot be directly interpreted as the chance probability of observing, from background alone, a positive fluctuation as large (or larger) as the observed one.
In order to perform the comparison with the other two methods we calibrated this $p_0$ probability with extensive simulations ($10^6$ realizations) of the null flux case. 
Namely, we compute the cumulative distribution of $p_0$ (see Fig.~\ref{fig:bayescalib}).
We note that such procedure is not purely Bayesian, but mixed with frequentist methodology.

Notably in the investigated toy MC case frequentist and Bayesian methods provide similar chance probability values (down to $\sim 10^{-5}$) at a given flux value. 
As expected, the agnostic method is less powerful.

\section{GW follow-up with IACT}\label{sec:gw}
\begin{figure}
    \centering
    \includegraphics[width=0.99\linewidth]{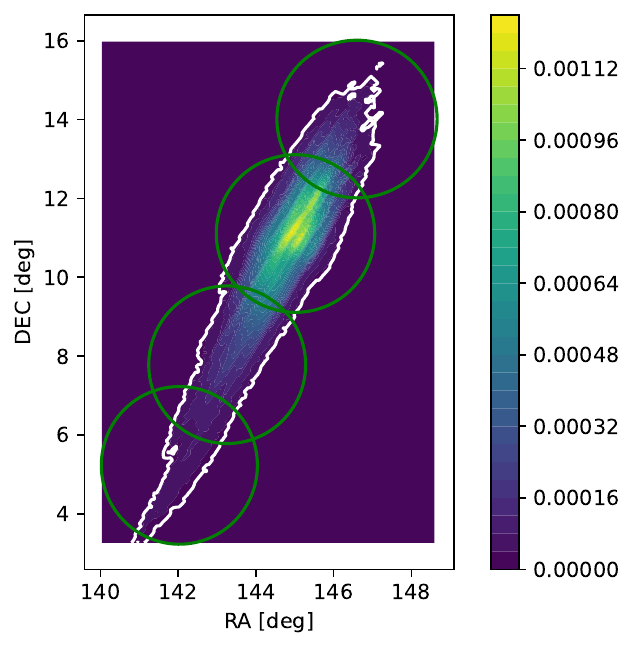}
    \caption{Probability map of the S250328ae GW alert in equatorial coordinates. White contour shows 95\% containment. Four green circles show hypothetical pointings with $2^\circ$ radius .}
    \label{fig:gw_prob}
\end{figure}

In order to test the methods in a realistic case we consider a hypothetical follow-up of GW candidate S250328ae \citep{2025GCN.39898....1L} with an array of four Large-Sized Telescopes (LSTs)  \citep{2024hegr.confE..26C}. 
Such telescopes are planned to form a part of the future Cherenkov Telescope Array Observatory \citep{2026NIMPA108771414O}.
The optimal pointing positions were determined using the \texttt{tilepy} software \citep{2026arXiv260118668S} assuming $2^\circ$ effective radius of field of view (FoV) of each observation.
The derived pointings, overlaid with the probability map of the alert, are shown in Fig.~\ref{fig:gw_prob}.
As the region of interest, ROI, we consider sky positions that are both within the 95\% containment region of the GW alert and within the FoV radius of $2^\circ$. 
The sky is binned in $0.07^\circ\times0.07^\circ$ bins, resulting in 4010 possible bins in such a region of interest, where a possible source can be present.
Because of the PSF of the instrument the signal from a source in one of those bins can contribute also to bins outside of ROI. 
Therefore in the calculations we consider event statistics in all 9605 bins (all bins with at least one pointing closer than $2^\circ$.
We use the instrument response functions (IRF) of the four LSTs array at zenith distance angle of $20^\circ$ averaged over North and South azimuth directions \citep{ctaprod5irf}.
We use the original binning of these IRFs, i.e. 0.1\,dex in true energy and 0.2\,dex in estimated energy.

At each true energy bin, using the IRF point-spread function, we calculate the position migration of the events (fraction of signal in i-th bin of the skymap if the source is in j-th bin). 
The response is averaged over the area of the i-th pixel. 
However, for simplicity we are neglecting a minor effect due to the orientation of the square pixel with respect to the direction towards the source measured on the camera plane. 
Assuming that each of the four pointings of the LST array lasts 20\,min, and background oversampling by a factor of 5, we prepare a background model using these IRFs. 
For simplicity and for linking the expected performance to plausible source fluxes, we convert the collection area to event rate from a source\footnote{This formula describes the power-law approximation of the of Crab Nebula, the standard candle of TeV astronomy, at the energies around 1 TeV\citep{2004ApJ...614..897A}} with flux $\mathrm{F.U.} \equiv dN/dE=2.83\times10^{-11} (E/\mathrm{TeV})^{-2.62}\mathrm{[TeV^{-1}s^{-1}cm^{-2}]}$.
With such a source model we also compute the energy migration between the true energy bins and estimated energy bins.

The application of the methods is nearly straightforward, as described in Section~\ref{sec:met}.
All the calculations are performed independently in each bin of estimated energy. 
Namely, in the agnostic approach, in each estimated energy bin, flux upper limit is computed as the least constraining limit in all the positions of the sky inside the ROI in that energy bin. 
The corresponding event statistics are integrated in the neighbouring bins with the angular cut corresponding to 75\% containment radius of the PSF at this energy. 
Similarly, in the frequentist approach we compute the TS(x) in each estimated energy bin according to Eq.~\ref{eq:freq}, using the same integration radius as in the agnostic method. 
Next TS(x) is maximized over $x$ and compared with the TS distribution in this energy bin for different fluxes. 
Finally, in the Bayesian approach we apply Eq.~\ref{eq:pbayes} in each estimated energy bin.
The expected signal $\mu_{i_{est}}$ in this estimated energy bin $i_{est}$ is computed by taking into account the energy resolution of the instrument and is being calculated by summation over all the true energy bins:
\begin{equation}
    \mu_{i_{est}}=\sum_{i_{true}}A_{\text{\it{eff}}, i_{true}}t_{\text{\it{eff}}} F_{i_{true}} M_{i_{true},i_{est}},
\end{equation}
where $A_{\text{\it{eff}}, i_{true}}t_{\text{\it{eff}}}$ is the exposure in this true energy bin, $F_{i_{true}}$ is the integrated flux over this true energy bin, and $M_{i_{true},i_{est}}$ is the energy migration matrix from $i_{true}$ true energy bin to $i_{est}$ estimated energy bin. 
Notably, both the exposure and the migration matrix depend on the position in the sky.
In this way, the flux constraints can be obtained independently in each estimated energy bin. 
As long as the assumed flux spectral shape agrees with the true one, or if in a particular energy bin there is not a large bias due to energy migration, such constraints are also good estimates of the constraints in the true energy bin. 

\begin{figure}
    \centering
    \includegraphics[width=0.49 \textwidth]{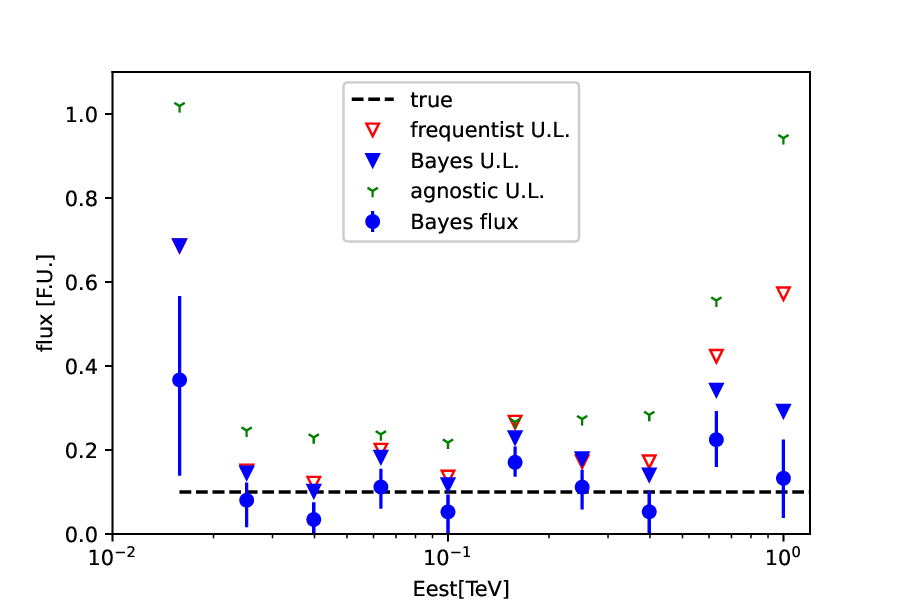}
    \caption{Comparison of flux reconstruction of different methods for the full simulations of a single realization example with a flux of 0.1. F.U. (horizontal dashed line) as a function of the estimated energy bin.
    Green tri-line symbol shown agnostic limit on the flux, empty red triangles show the frequentist limit. Blue filled markers show the Bayesian flux estimate (circle) and upper limit (triangle). 
    }
    \label{fig:tilepyex}
\end{figure}

In Fig.~\ref{fig:tilepyex} we show the example calculations for the simulated flux of 0.1~F.U..
Due to the power-law nature of the source model, and the rapidly dropping residual background vs. energy, different energies scan a wide range of  event statistics, with $\sim 90$ background events per skymap bin at 25 GeV down to $0.3$ at 1~TeV. 
In the presented example, all three models provide 95\% C.L. upper limits that are above the simulated flux. 
As expected in most of the energy bins the agnostic limits are the least constraining. 

We compare the performance of the methods by simulating 1000 realizations of the skymap (including also randomization of the alert position) for 40 logarithmically spaced true flux assumptions between 0.01 and 0.96~F.U..
The realizations occurring outside of the defined ROI are discarded.
To facilitate numerical computations, comparing to the example from Fig.~\ref{fig:tilepyex}, we apply two minor simplifications.
First, we only consider true positions of the signals in the middle of a bin. 
Second, in the frequentist method, 
the obtained best TS must be compared to the library of MC with various values of the true flux. 
Instead of generating a separate library for each of the 1000 realizations with the background measured from this particular realization, we generate only a single library with average background. 

\begin{figure*}
    \includegraphics[width=0.99\textwidth]{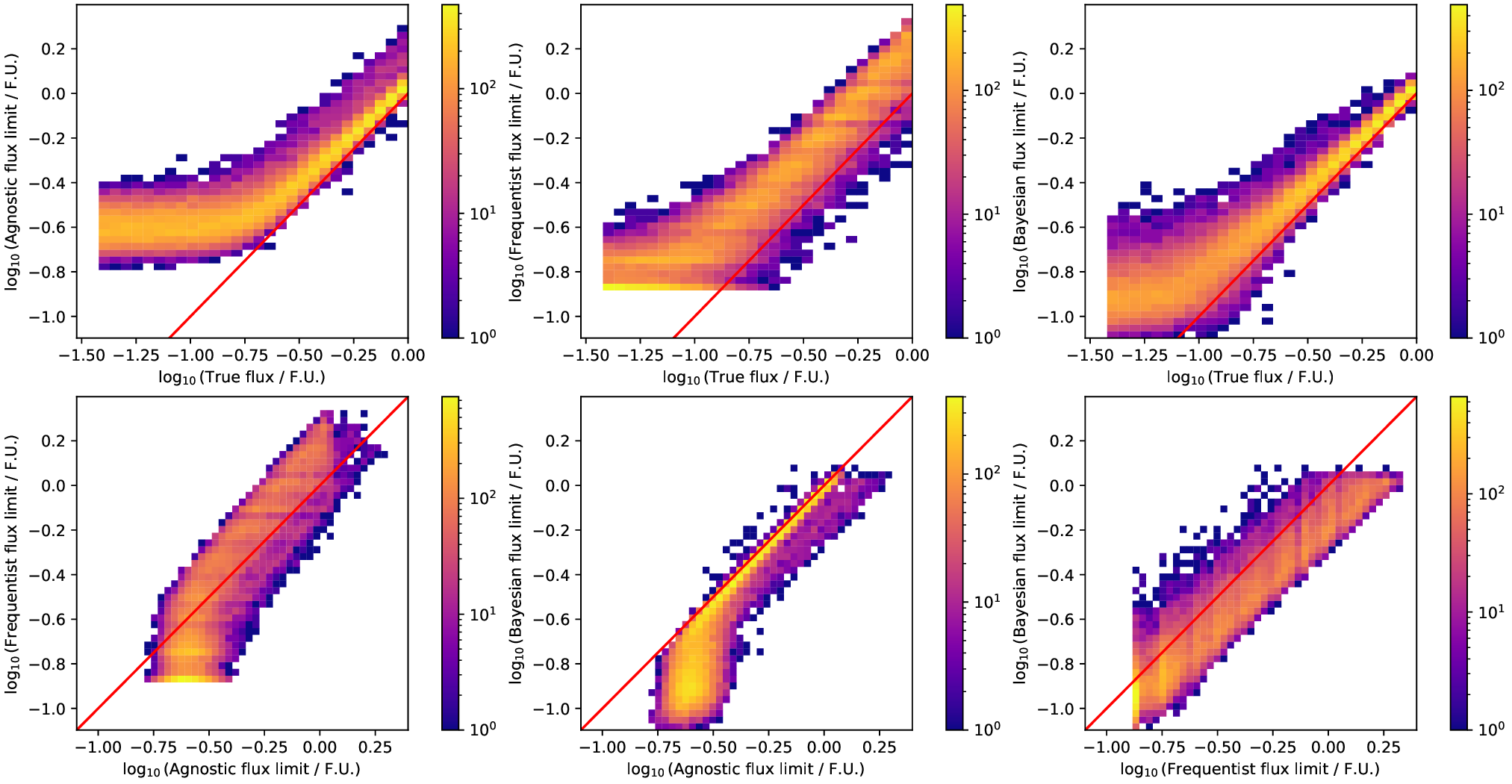}
    \caption{Comparison of the limits obtained with three methods (as in Fig.~\ref{fig:toycomp}) for full simulations at 63 GeV. 
    }
    \label{fig:comp_tilepy}
\end{figure*}

The results of the comparisons of the three methods are shown in Fig.~\ref{fig:comp_tilepy} for the example bin centered at the estimated energy of 63 GeV.
The full simulations reproduce the basic features visible in the case of the toy MC study.
As expected, the agnostic method is the least powerful. 
The frequentist method supersedes it, especially for the lowest simulated fluxes.
The Bayesian method provides nominally, on average, a slightly lower limit than frequentist approach. 
As we increase the injected true flux, the distributions of the limits from the agnostic and the Bayesian methods become narrower and peak just above the true flux. The limits from the frequentist method in that high-signal regime are on average higher, and have a larger spread. This is the same effect seen in the one-dimensional test: the frequentist approach does not make use of the positional information provided by the strong gamma signal, and keeps simulating source locations according to the GW probability map (hence exploring regions of widely different gamma acceptance), which results in less constraining limits - of course, in such a case one could perform a regular flux estimate at the position of the observed excess.

Contrary to the toy MC case for a strong signal the Bayesian limit does not converge exactly on the value of the agnostic limit. 
This is likely related to the treatment of the PSF of the instrument. 
The agnostic limit simply integrated the number of events within the angular cut around the putative source position. 
In contrast, the Bayesian approach is using in the likelihood the shape of the PSF of the instrument.
On one hand using such additional information is likely to improve the performance, and thus lower the value of the limit.
However, it also makes the method slightly more affected by the difference of the true source position (that can be arbitrary) and the scanned positions (which are only in the centres of the skymap bins). 

\begin{figure}
    \centering
    \includegraphics[width=0.49\textwidth]{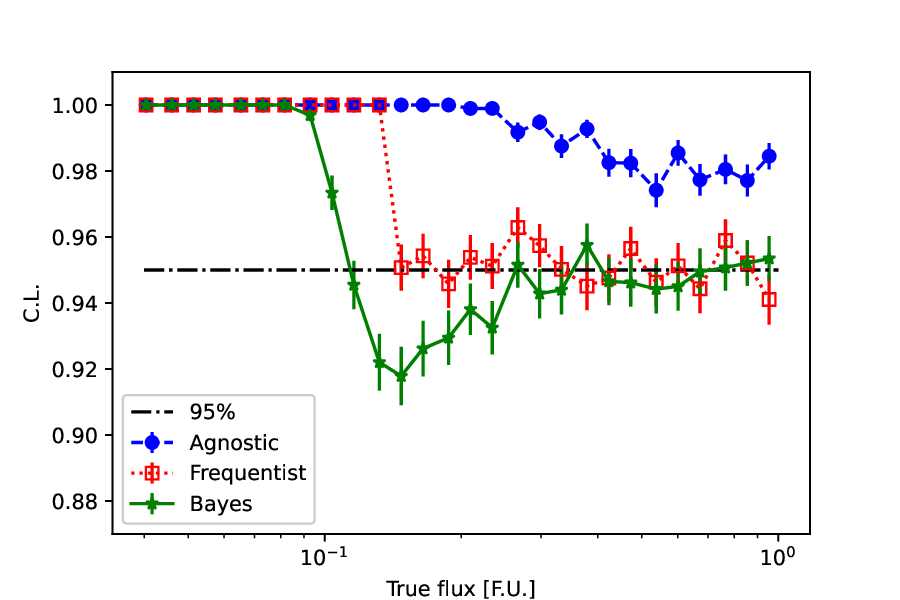}
    \caption{Fraction of realization with a given true flux that result in the limit above the true flux value for the three methods: agnostic (blue filled dots and dashed line), frequentist (red empty square and dotted line) and Bayesian (green asterisks and solid line). 95\% is shown with gray dot-dashed line. Full simulations at the energy of 63 GeV.}
    \label{fig:tilepycl}
\end{figure}

In Fig.~\ref{fig:tilepycl} we show the C.L. of the limits achieved with all the investigated methods. 
We evaluate it by computing the fraction of the simulated realization with the derived limit exceeding the simulated flux. 
While in principle it is not expected from the Bayesian method to reproduce 95\% C.L. in such frequentist-style test, the obtained C.L. from MC simulations is also close to 95\% in this method.
The general behaviour of the empirically derived C.L. for those full simulations closely follow the behaviour in the toy MC simulations described in Section~\ref{sec:toy}.

\begin{figure}
    \centering
    \includegraphics[width=0.49\textwidth]{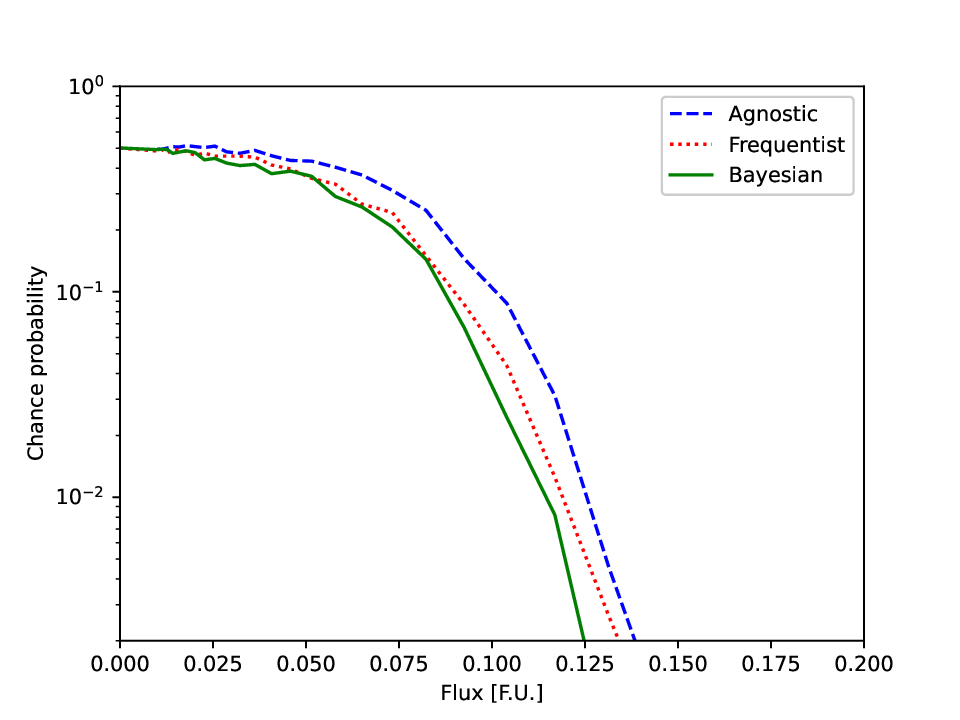}
    \caption{The chance probability of observing, from background alone, a fluctuation at least as large as the observed excess in agnostic (blue dashed), frequentist (red dotted) and Bayesian (green solid) approach using full simulations at the energy of 63 GeV. }
    \label{fig:detection_tilepy}
\end{figure}

Similarly to Section~\ref{sec:detection} we evaluate the chance probability of reaching a particular detection level. 
The chance probabilities at the example energy of 63~GeV are shown in Fig.~\ref{fig:detection_tilepy}.
Similarly to the toy MC study both frequentist and Bayesian approach lead a similar sensitivity for detection.
The sensitivity of the agnostic method is also similarly degraded. 

We have not considered instrumental systematic uncertainties (e.g. in the signal efficiency) in any of the calculations. In the case of the agnostic method, they are trivial to include following the approach of  \citet{2005NIMPA.551..493R}. Similarly, it is also easy to include them in the frequentist approach, by randomizing (using a particular model of systematic uncertainties) the individual realizations in the MC library used in the comparison with the data. The Bayesian approach also provides a natural way of including the systematic uncertainties by adding a nuisance parameter with a prior distribution reflecting our constraints on them. This requires marginalization of the probability over one additional dimension. An example of introduction of systematic uncertainties in the toy MC case is presented in Appendix~\ref{sec:syst}.

\section{Discussion and conclusions}\label{sec:conc}
We studied three types of methods used to derive limits on the emission from a source, whose position is not well known. 
The first method is agnostic on the probability distribution of the position of the source (but still assumes a ROI on the sky from which the emission might originate). 
The other two methods exploit this information, either in frequentist or Bayesian approach. 
The numerical code exploiting those methods is made available in \url{https://github.com/jsitarek/uncertain_position_limits}.
We investigated the basic features of all the three methods with a toy MC case, and next we applied them to full simulations of a realistic follow up of a GW event with a sub-array of four LST telescopes. 
As expected, the inclusion of the information about the source position probability distribution improves the achievable constraints on the source flux. 
Both frequentist and Bayesian methods provide a similar improvement in the derived limits on the emission.
Both methods present minor technical challenges.
Namely, the frequentist approach requires a generation of a library of MC realization with different levels of the emission, and the Bayesian approach requires a numerical marginalization over source position. 
Nevertheless, we show that such calculations are numerically feasible in realistic case of a GW alert follow-up with Cherenkov telescopes even if the scanned region is large enough to require multiple pointings. 
The presented methods are also naturally extendable to the inclusion of the effect of various systematic uncertainties. 
We note that the methods are generic and can be applied to various types of alerts (such as e.g. gravitational waves, poorly localized neutrino events, gamma ray bursts with uncertain localization) with various types of instruments (not only Cherenkov telescopes, but also satellite detectors, ground based gamma-ray surface arrays, or even cosmic ray or neutrino detectors). 
The presented implementation is particularly efficient for instruments with high event rates in which binned analysis is preferred.

\section*{Acknowledgements}
\includegraphics[width=0.4 \textwidth]{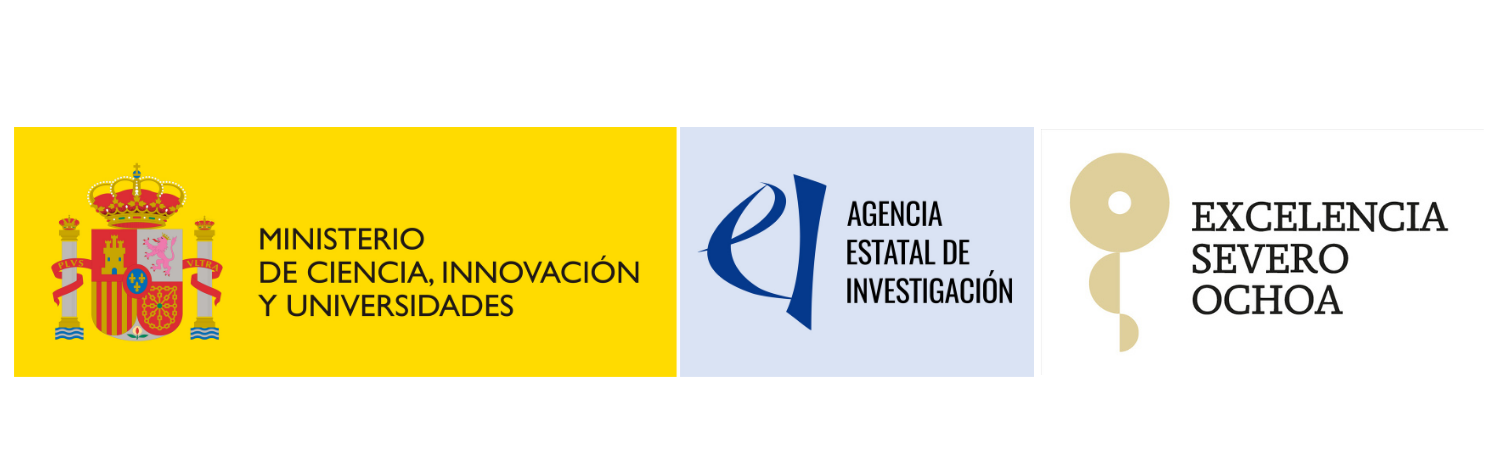}

The project BPN/BIT/2024/1/00037/U/00001 is co-fi\-nan\-ced by the Polish National Agency for Academic Exchange and MAECI: PL25MO12 (Polish-Italian ``Canaletto'' programme).
JS would like to thank IFAE for hosting him for a stay funded by the grant «Excelencia Severo Ochoa» CEX2024-001441-S by 
MICIU/AEI/10.13039/501100011033.
G.D.’s work on this project was supported by the Beatriu de Pinós programme (BP 2023).
Authors would like to thank M. Seglar for discussions about tilepy and anonymous journal reviewer for comments that helped to improve the manuscript.

\appendix
\section{Region of interest in Bayesian method}
\label{sec:singlebin}
It is tempting to simplify the calculation of the total probability for a position $x$ in the Bayesian method by including in Eq.~\ref{eq:pbayes} only the bins $x'$ that are close to $x$ (i.e. in which $PSF(x,x')$ is not close to 0). 
Those bins are expected to bring most of the information about the plausibility of the occurrence of a source at the position $x$ with a particular flux. 
This however brings two problems. 

First, the data used in the estimation of probabilities at different $x$ vary, hence the normalization of the sum $p(x,f)$ to 1 is even more arbitrary.
As a result, $p(x,f)$ loses its strict Bayesian meaning of posterior distribution given the assumed priors and the observations. 
It is not possible to illustrate this with the toy MC case presented in Section~\ref{sec:toy}, due to the negligible probability of reproducing exactly the same bin-wise statistics with a random realization.
Thus, we validated this in the even more simplified case of 2 position bins, 4 flux states and perfectly known background estimate. 
Namely, for a particular realization of the event statistics in the 2 bins, with a given prior, we constructed the Bayesian posterior distribution either using directly Eq.~\ref{eq:pbayes},  or limiting it to just the probability in one bin, the one in which the presence of a signal is assumed. 
We then construct the MC posterior distribution by generating a large number of random realizations with the given priors and picking the position and flux parameters of those that result in the same event statistics in both bins as the first realization. 
As expected from Bayes theorem, the posterior computed using Eq.~\ref{eq:pbayes} matches perfectly the empirical MC posterior. 
In contrast, neglecting non-signal bins in this equation produces a clear mismatch between them.
However the actual difference depends on the specific realization and details of the test, thus such simplified toy calculations cannot be  generalized either to the toy MC case or to the full simulation case considered in the main part of the paper. 

\begin{figure*}
    \centering
    \includegraphics[width=0.32\linewidth]{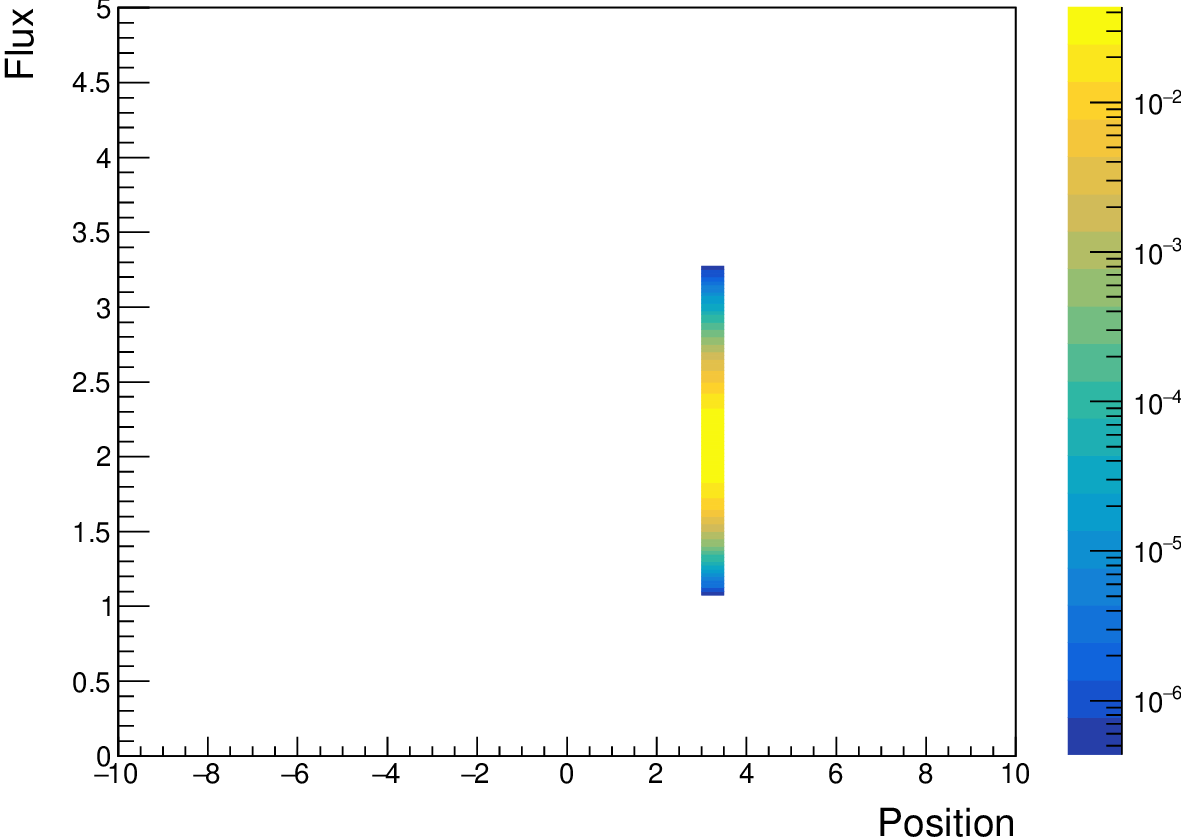}
    \includegraphics[width=0.32\linewidth]{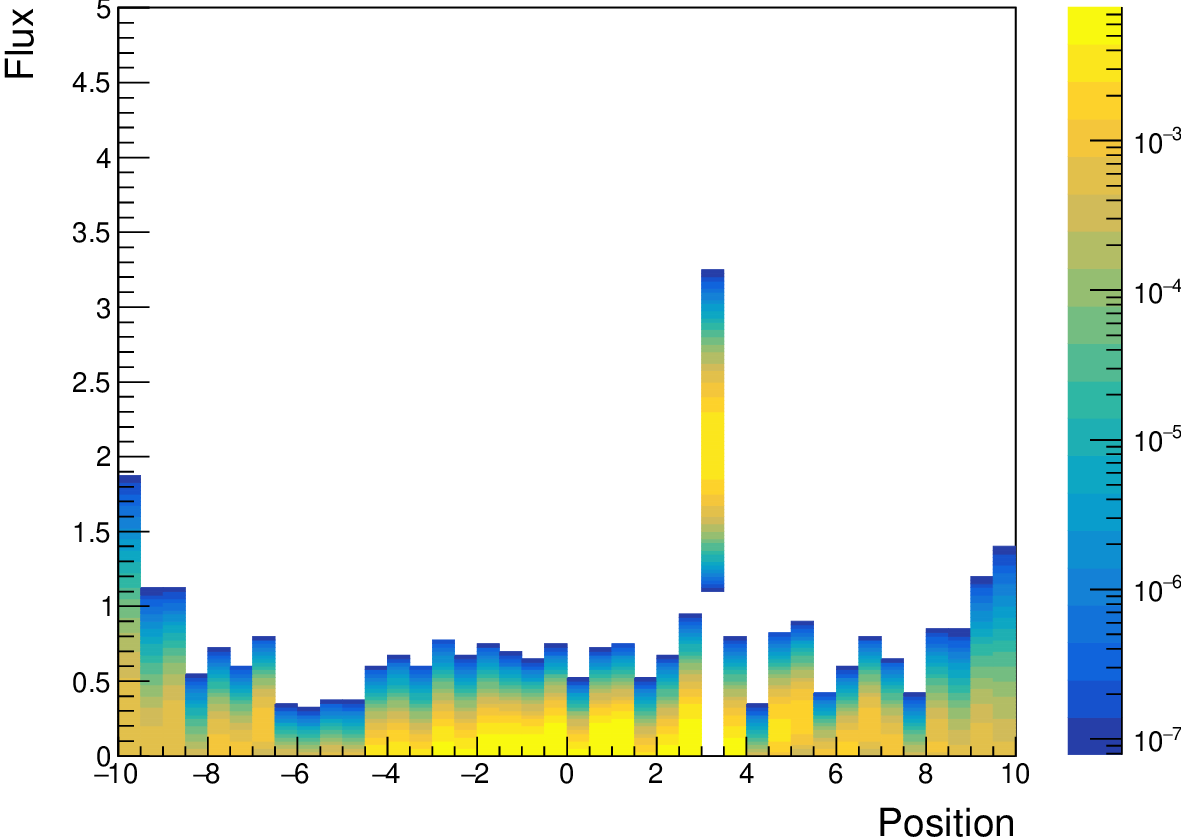}
    \includegraphics[width=0.32\linewidth]{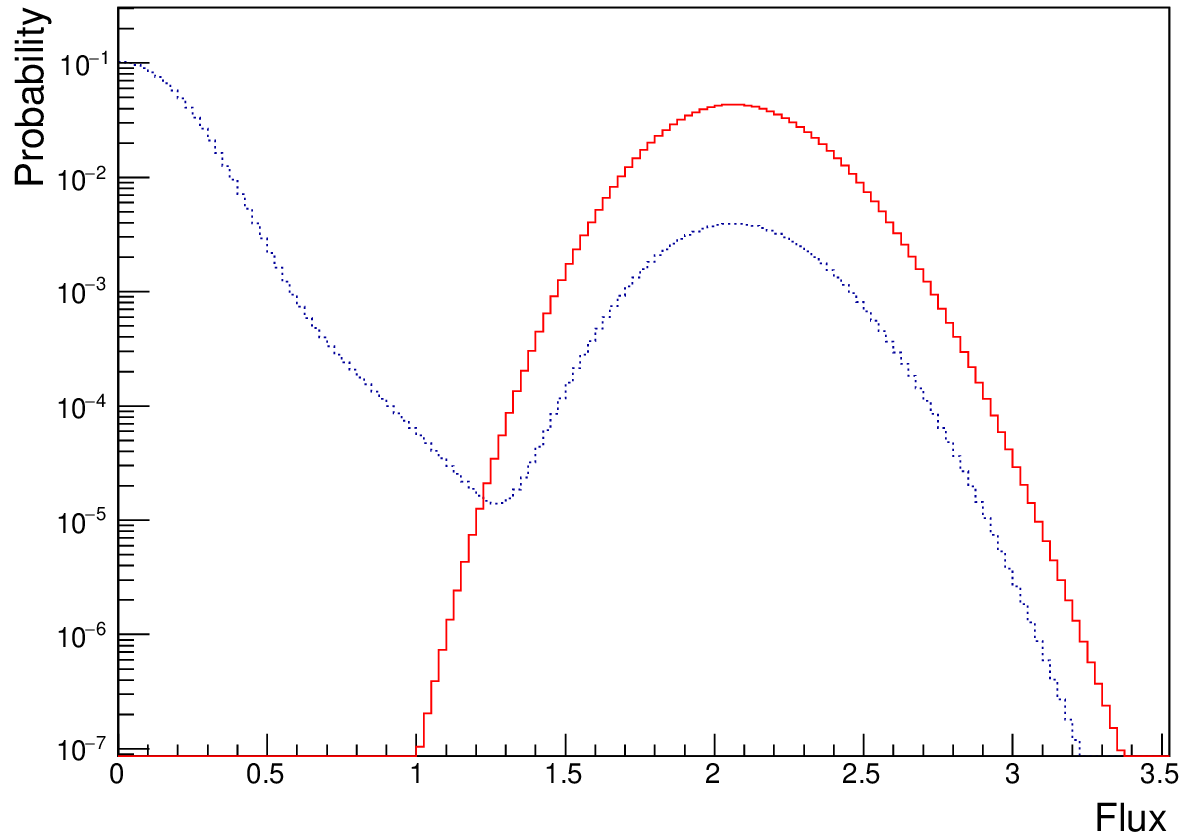}
    \caption{Comparison of posterior probability distribution on the flux and position for the calculations according to Eq.~\ref{eq:pbayes} (left panel) and when limiting only to the probability of the signal bin (central panel).
    The probabilities at least $10^5$ times smaller than the peak probability are plotted in white.
    Right panel shows the flux distribution probability marginalized over position from the left panel (red solid curve) and from the middle panel (black dotted). True flux was set to a value of 2. }
    \label{fig:bayes_singlebin}
\end{figure*}

Second, neglecting non-signal bins has also an effect on estimation of the probability for fluxes producing significant emission. 
In Fig.~\ref{fig:bayes_singlebin} we present a comparison of using all the bins in the calculation of the Bayesian probability and limiting it only to the bin of the assumed source position.
Considering all the bins in the probability according to Eq.~\ref{eq:pbayes} strongly suppresses the posterior probability of having a weak signal even in the positions not consistent with the true signal. 
On the other hand, limiting the posterior probability calculation to only the bin of the corresponding signal provides large probability for weak signal in all the bins except the one with true signal.
The peak probability is actually for zero flux (which is expected because the probability corresponding to the most probable value of a Poissonian distribution decreases as its expected value increases). 
The effect of the null-flux probabilities gets additionally enhanced by a factor of the number of position bins when marginalizing over the position. 
As a result neglecting non-signal bins in the Bayesian approach would result in heavily biased posterior flux probability distribution in case of a significant flux. 

\section{Confidence level of non-negative flux}\label{sec:cl_f0}
As discussed in Section~\ref{sec:freq} in the frequentist approach (contrary to the Bayesian one), there is no natural way of introducing a condition for non-negative flux, which could result in non-physical negative flux upper limits of a given C.L. (see also the discussion in \citealp{2005NIMPA.551..493R}). 
Therefore, we used in the frequentist approach clipping of the negative fluctuations of the test statistics at the median value obtained from the null flux simulations.
Here we propose an alternative way of presenting limits for such cases. Instead of computing the flux upper limit at a given C.L. we calculate the C.L. for the flux to be positive. 
In the frequentist approach this is straightforward by checking the fraction of simulations with null flux that result in a TS value below the TS value of the investigated realization (without clipping).
In the Bayesian approach we use a similar procedure as in investigating the significance of the detection, namely we artificially allow negative fluxes and integrate the posterior probability distribution above zero flux. 
We compare the two methods using the toy MC setup described in Section~\ref{sec:toy}.

\begin{figure*}
    \centering
    \includegraphics[width=0.4\textwidth]{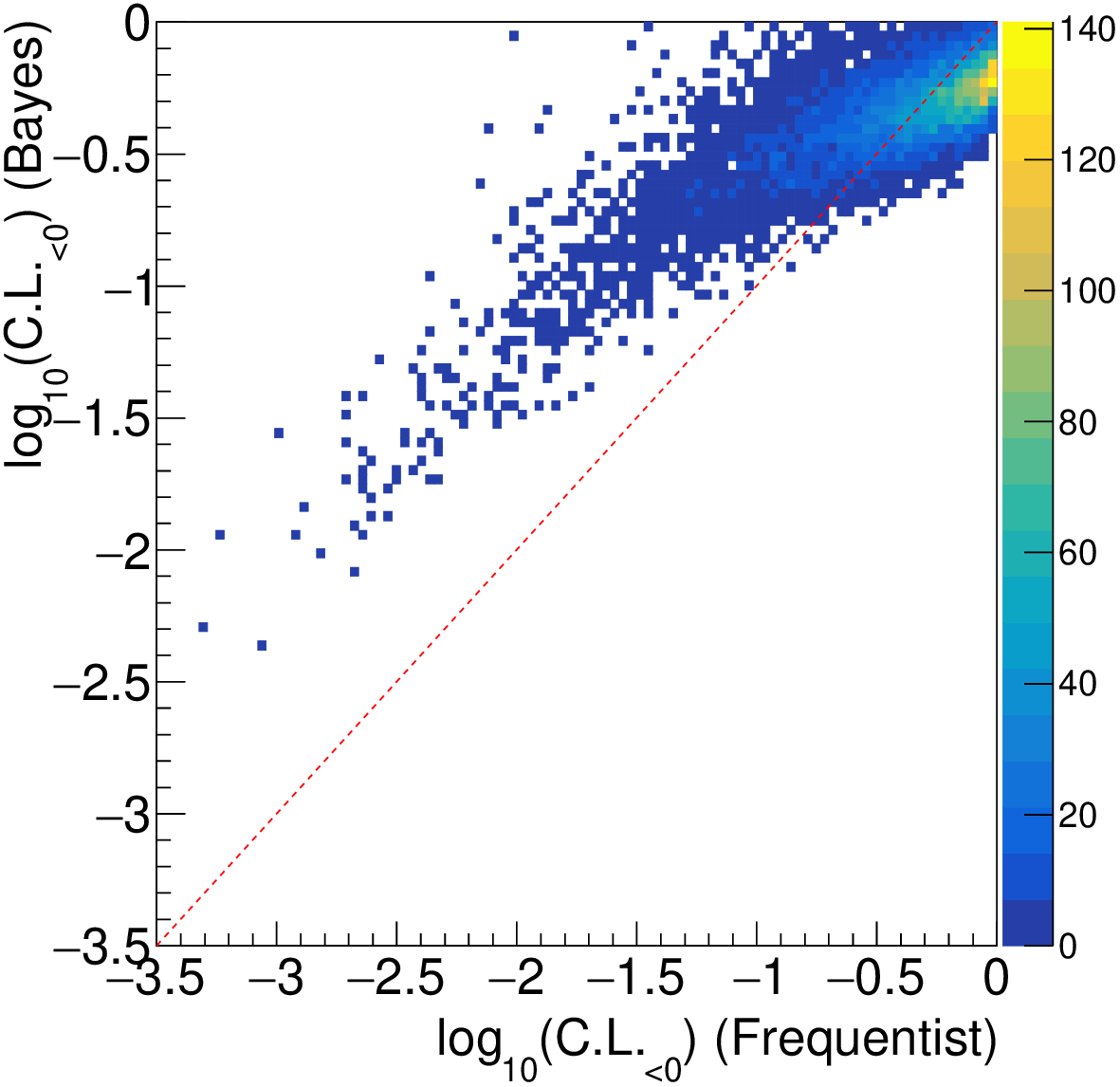}
    \hspace{1cm}
    \includegraphics[width=0.4\textwidth]{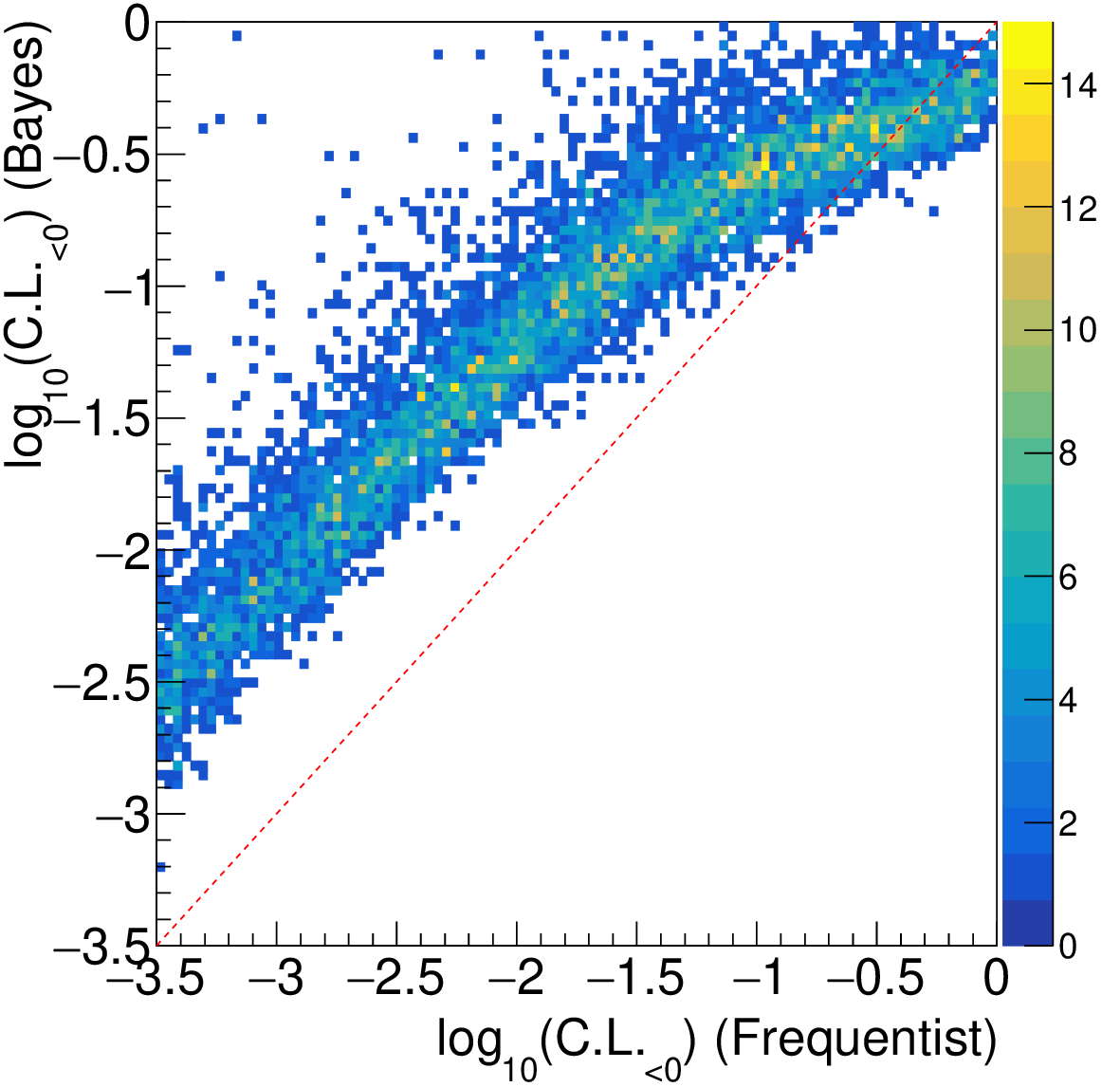}
    \caption{Comparison of the residual confidence level on ``negative'' flux of Bayesian vs frequentist approach.
    Toy MC setup with null flux (left panel) and flux of 0.5 a.u. (right panel).
    }
    \label{fig:cl_f0}
\end{figure*}

In Fig.~\ref{fig:cl_f0} we present the complement of such probability to 1 (i.e. integrated probability of explaining the observations with a negative flux). 
The value of probability are in general higher for the Bayesian approach, which might make an impression that the frequentist approach is more efficient at rejecting null (or negative) flux hypothesis. 
It should be noted, however, that by extending the probability distribution to negative fluxes the power of Bayesian method is artificially reduced (by removing a natural constraint on the posterior).
Moreover, as discussed in Section~\ref{sec:detection} for the null flux case the posterior probability integrated for positive fluxes does not have a flat distribution.
We insist, however, in the fact that taking advantage of negative fluctuations in an observation (e.g. when obtaining in the frequentist method a TS significantly below the null-hypothesis median TS) to set more constraining limits, has a risk: should the observation result from hidden systematics rather than from a purely statistical fluctuation, the reported limit could fail to have the intended coverage.

\section{Inclusion of systematic uncertainties in Bayesian approach} \label{sec:syst}
It is straightforward to include a systematic uncertainty in the reconstructed flux of the instrument in the calculations of the Bayesian method.
Such uncertainty can be introduced as a nuisance parameter with a prior representing our knowledge of the possible systematic errors of the instrument.
The systematic uncertainties are instrument-specific, and might also depend on observation conditions and particular type of source. 
For example \textit{Fermi}-LAT considers 5-15\% uncertainty in the flux depending on the PSF type of the event \citep{2020ApJ...892..105A}.
For IACTs, the total systematic uncertainty in the flux (that includes also uncertainty in the energy scale) is larger, and typically quoted as $\sim 30\%$ (see e.g. \citealp{jsgg2024}).

\begin{figure*}
    \centering
    \includegraphics[width=0.95\textwidth]{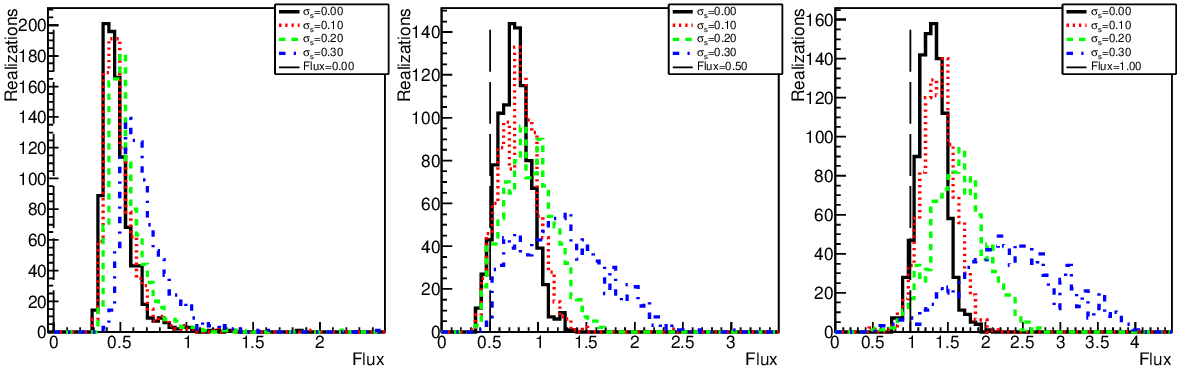}
    \caption{Distribution of flux upper limits obtained in the toy MC case with inclusion of systematics with a Gaussian standard deviation of 0 (black solid), 10\% (red dotted), 20\% (green dashed) and 30\% (blue dot-dashed). 
    The simulated flux (vertical long-dashed black line) is 0 (left panel) 0.5 (middle) and 1 (right). }
    \label{fig:bayessyst}
\end{figure*}

Including the systematic uncertainty $s$ on the flux (which is roughly equivalent to the systematic uncertainty on the collection area) as a nuisance parameter requires a modification of Eq.~\ref{eq:pbayes} to:
\begin{eqnarray}
    p(x, f, s)&=&A\prod_{x'}P(N_{ON}(x'), \mathrm{PSF}(x,x')s\mu+b(x')) \nonumber \\
    &\times& P(N_{OFF}(x'), b(x')\beta) p_{GW}(x') p_s(s).  \label{eq:pbayessyst}  
\end{eqnarray}
In order to obtain $p_f(f)$, in comparison to Eq.~\ref{eq:pf} additional integration/summation over $s$ is required. 
We assume a Gaussian prior on $s$: 
\begin{equation}
    p_s(s)=\exp\left(-(s-1)^2/(2\sigma_s^2)\right)/\sqrt{2\pi\sigma_s^2},
\end{equation}
clipped to the $[1-2\sigma_s, 1+2\sigma_s]$ range.
We consider four values of $\sigma_s=0, 10\%, 20\%, 30\%$ and apply it in the toy MC case, both varying the simulated event statistics with the systematic uncertainty and using  Eq.~\ref{eq:pbayessyst} in the derivation of the Bayesian upper limit on the flux. 
In Fig.~\ref{fig:bayessyst} we present the distribution of the obtained limits for three considered fluxes: 0, 0.5 and 1. 
As expected, inclusion of systematic uncertainties affects the achieved performance by making the distribution of the obtained limits broader and shifted to higher values. 
The effect becomes large for strong fluxes and 30\% systematic uncertainties.
This is due to a possibility of ``hiding'' a strong flux in the position of low exposure combined with non negligible probability of low $s$ value.

\end{document}